\documentstyle[12pt,epsf]{article}
 \newcommand{\insertplot}[5]{\begin{figure}
 \hfill\hbox to 0.05in{\vbox to #5in{\vfill
 \inputplot{#1}{#4}{#5}}\hfill}
 \hfill\vspace{-.1in}
 \caption{#2}\label{#3}
 \end{figure}}
 \newcommand{\inputplot}[3]{
 \special{ps: plotfile #1}
}

\begin{document}
 
\title{Dyonic Non-Abelian Black Holes}
\vspace{1.5truecm}
\author{
{\bf Yves Brihaye$^1$, Betti Hartmann$^2$,
Jutta Kunz$^2$ and Nad\`ege Tell$^1$}\\
$^1$ Facult\'e des Sciences, Universit\'e de Mons-Hainaut\\
B-7000 Mons, Belgium\\
$^2$Fachbereich Physik, Universit\"at Oldenburg, Postfach 2503\\
D-26111 Oldenburg, Germany}
 

\maketitle
\vspace{1.0truecm}

\begin{abstract}
We study static spherically symmetric 
dyonic black holes in Einstein-Yang-Mills-Higgs theory.
As for the magnetic non-abelian black holes,
the domain of existence of the dyonic non-abelian black holes 
is limited with respect to the horizon radius and the
dimensionless coupling constant $\alpha$,
which is proportional to the ratio of vector meson mass and Planck mass.
At a certain critical value of this coupling constant, $\hat \alpha$,
the maximal horizon radius is attained.
We derive analytically a relation between $\hat \alpha$
and the charge of the black hole solutions
and confirm this relation numerically.
Besides the fundamental dyonic non-abelian
black holes, we study radially excited dyonic non-abelian
black holes and globally regular gravitating dyons.
\end{abstract}
\vfill
\noindent {Preprint hep-th/9904065} \hfill\break
\vfill\eject

\section{Introduction}

SU(2) Yang-Mills-Higgs (YMH) theory, with the Higgs field
in the adjoint representation,
possesses globally regular particle-like solutions,
such as the
't Hooft-Polyakov magnetic monopole \cite{thooft}
and the Julia-Zee dyon, 
which carries both magnetic and electric charge \cite{jul}.

In SU(2) Einstein-Yang-Mills-Higgs (EYMH) theory,
for small gravitational constant
the gravitating fundamental monopole
\cite{old,ewein,bfm}
and the gravitating fundamental dyon \cite{bhk}
emerge smoothly
from the corresponding flat space solutions.
The mass of the gravitating fundamental monopole \cite{ewein,bfm}
and dyon \cite{bhk} solutions decreases
with increasing gravitational constant,
corresponding to increasing coupling constant $\alpha$,
which is proportional to the ratio of vector meson mass and Planck mass.
The regular solutions cease to exist beyond some maximal value
of $\alpha$, depending on the electric charge $Q$ 
\cite{ewein,bfm,bhk,ewein3}.

Besides these globally regular solutions, in SU(2) EYMH theory
magnetic
\cite{ewein,bfm,bfm2,aichel} and dyonic \cite{bhk}
non-abelian black hole solutions exist.
They emerge from the globally regular solutions,
when a finite regular event horizon $x_h$ is imposed. 
Consequently,
they have been characterized as ``black holes within magnetic monopoles''
\cite{ewein} or ``black holes within dyons'' \cite{bhk}.
Distinct from the corresponding embedded Reissner-Nordstr\o m (RN) 
black holes \cite{rn},
these non-abelian black hole solutions
represent counterexamples to the ``no-hair'' conjecture,
because they carry non-trivial non-Coulomb-like fields
outside their horizon.
In contrast, pure
SU(2) EYM theory possesses neither regular dyon solutions \cite{gal1},
nor dyonic black holes other than embedded RN solutions \cite{gal2,popp}.

The magnetic non-abelian black holes
exist only in a limited domain of the $\alpha-x_h$ plane 
\cite{ewein,bfm,bfm2}.
The domain of existence consists of two regimes,
where for fixed coupling constant $\alpha$ and varying 
horizon radius $x_h$,
the solutions approach limiting solutions in two distinct ways.
At the transition point between these two regimes, given by
the critical value $\hat \alpha$, the maximal horizon radius
is attained \cite{bfm,bfm2}.
Limiting our investigations to vanishing Higgs self-coupling,
we here determine the domain of existence 
of the dyonic non-abelian black holes,
finding also two regimes for the coupling constant $\alpha$.

Besides the fundamental gravitating monopole and dyon solutions
there are radially excited solutions \cite{bfm,bfm2,bhk},
where the gauge field function possesses nodes.
Both the radially excited monopole and dyon solutions 
are related to the globally regular
Einstein-Yang-Mills (EYM) solutions, 
found by Bartnik and McKinnon \cite{bm},
and, like these solutions, have no flat space counterparts.
Besides the radially excited globally regular dyon solutions,
we here consider the radially excited dyonic non-abelian
black hole solutions and determine their domain of existence.

In section II we derive the EYMH equations of motion.
We present the equations both in Schwarzschild-like coordinates,
employed in the numerical calculations \cite{bhk},
and in isotropic coordinates, to extend the analytical calculations
\cite{bfm2} to the dyonic case.
In section III we briefly review the embedded RN solutions.
We discuss the globally regular dyon solutions in section IV.
In addition to vanishing Higgs self-coupling,
we consider the fundamental solutions also for finite Higgs self-coupling,
in order to demonstrate that, as in flat space \cite{bkt},
a maximally possible charge $Q_{\rm max}$ arises.
In section V we discuss the dyonic non-abelian black hole solutions.
In particular, we derive the critical value $\hat \alpha$ 
for the dyonic solutions analytically and confirm it numerically.
For vanishing Higgs self-coupling,
we determine the domain of existence for the fundamental
and the first radially excited dyonic non-abelian black hole solutions,
and we discuss the pattern how the black hole solutions
approach limiting solutions.
We conclude in section VI.
In Appendix A, finally we show the local existence of the non-abelian
regular dyon and dyonic black hole solutions
at the origin and at the horizon, respectively.

\section{Einstein-Yang-Mills-Higgs Equations of Motion}

We consider the SU(2) EYMH action
\begin{equation}
S=S_G+S_M=\int L_G \sqrt{-\rm g} d^4x + \int L_M \sqrt{-\rm g} d^4x
\ \label{action}   \end{equation}
with
\begin{equation}
L_G=\frac{1}{16\pi G}{\cal R}
\ , \end{equation}
and
\begin{equation}
L_M = - \frac{1}{4} F_{\mu\nu}^a F^{a\mu\nu} 
      - \frac{1}{2} D_\mu \phi^a D^\mu \phi^a
      - \frac{\beta^2 g^2}{4} (\phi^a \phi^a - v^2)^2
\ , \end{equation}
where
\begin{equation}
F_{\mu\nu}^a = \partial_\mu A_\nu^a - \partial_\nu A_\mu^a
 + g \epsilon^{abc} A_\mu^b A_\nu^c
\ , \end{equation}
\begin{equation}
D_\mu \phi^a = \partial_\mu \phi^a + g \epsilon^{abc} A_\mu^b \phi^c
\ , \end{equation}
$g$ is the gauge coupling constant, $\beta$ is the dimensionless
Higgs coupling constant, proportional to the ratio of Higgs boson mass
and vector boson mass,
and $v$ is the Higgs field vacuum expectation value.
Variation of the action eq.~(\ref{action}) with respect to the metric
$g_{\mu\nu}$, the gauge field $A_\mu^a$ and the Higgs field $\phi^a$
leads to the Einstein equations and the matter field equations.
 
\subsection{Ansatz}

For static spherically symmetric globally regular
and black hole solutions the metric can be parametrized as \cite{bfm2,bck}
\begin{equation}
ds^2=g_{\mu\nu}dx^\mu dx^\nu=
  - e^{2 \nu (R)} dt^2 + e^{2 \lambda (R)} dR^2 
  + r^2(R) (d\theta^2 + \sin^2\theta d\phi^2)
\ . \label{genmet} \end{equation}

For the gauge and Higgs field we employ the
spherically symmetric ansatz \cite{bhk}
\begin{equation}
\vec A_t = \vec e_r J(r) v
\ , \end{equation}
\begin{equation}
\vec A_r=0 \ , \ \ \
\vec A_\theta =  -\vec e_\phi \frac{1- K(r)}{g} \ , \ \ \
\vec A_\phi =   \vec e_\theta \frac{1- K(r)}{g} \sin \theta
\ , \end{equation}
and 
\begin{equation}
\vec \phi = \vec e_r H(r) v
\ , \end{equation}
with the standard unit vectors $\vec e_r$, $\vec e_\theta$ and $\vec e_\phi$.
For $A_t^a=0$, the solutions carry only magnetic charge
\cite{ewein,bfm,aichel}.

With these ans\"atze we obtain the action
\begin{eqnarray}
{S} = \frac{1}{G} \int dt dR e^{\nu+\lambda} && \Biggl(  
    {1\over 2} \left[1+e^{-2\lambda} \left((r')^2
  +\nu'(r^2)'\right)\right] \Biggr. \nonumber\\
 &&+4 \pi G \left[
      e^{-2(\lambda+\nu)} \frac{r^2}{2} (J')^2 v^2
    + e^{-2\nu} K^2J^2v^2 \right. \nonumber\\
 && \Biggl. \left.
    - e^{-2\lambda} \left( \frac{(K')^2}{g^2} 
    + {r^2\over 2} (H')^2v^2) \right) - V_2 \right] 
                         \Biggr)
\ , \label{ac1} \end{eqnarray}
where $\nu$, $\lambda$, $r$, $K$, $J$ and $H$ are functions of $R$,
the prime indicates the derivative with respect to $R$, and 
\begin{equation}
V_2 = {(1-K^2)^2\over{2g^2r^2}} + K^2 H^2 v^2
+ {\beta^2 g^2 \over 4} r^2 (H^2-1)^2 v^4
\ . \label{v2} \end{equation}

\subsection{Schwarzschild-like coordinates}

We first employ Schwarz\-schild-like coordinates,
corresponding to the gauge $r(R)=R$. Renaming $R=r$, 
the spherically symmetric metric reads
\begin{equation}
ds^2=
  -A^2N dt^2 + N^{-1} dr^2 + r^2 (d\theta^2 + \sin^2\theta d\phi^2)
\ , \end{equation}
with the metric functions 
\begin{equation}
A(r)= e^{\lambda + \nu}
\ , \label{a} \end{equation}
\begin{equation}
N(r)=1-\frac{2m(r)}{r} = e^{-2 \lambda}
\ , \label{n} \end{equation}
and the mass function $m(r)$.

We now introduce the dimensionless coordinate $x$
and the dimensionless mass function $\mu$,
\begin{equation}
x = g v r \ , \ \ \ \mu=g v m
\ , \label{xm} \end{equation}
as well as the dimensionless coupling constant $\alpha$
\begin{equation}
\alpha^2 = 4 \pi G v^2 
\ . \end{equation}

The $tt$ and $rr$ components of the Einstein equations then yield
the equations for the metric functions,
\begin{eqnarray}
\mu'&=&\alpha^2 \Biggl( \frac{x^2 J'^2}{2 A^2} 
   + \frac{J^2 K^2}{A^2 N}
   + N K'^2 + \frac{1}{2} N x^2 H'^2
\nonumber\\
 & &\phantom{ \alpha^2 \Biggl( }
   + \frac{(K^2-1)^2}{2 x^2} + H^2 K^2
   + \frac{\beta^2}{4} x^2 (H^2-1)^2 \Biggr)
\ , \label{eqmu} \end{eqnarray}
and
\begin{eqnarray}
 A'&=&\alpha^2 x \Biggl(
    \frac{2 J^2 K^2}{ A^2 N^2 x^2}
   + \frac{2 K'^2}{x^2} + H'^2 \Biggr) A
\ , \label{eqa} \end{eqnarray}
where the prime now indicates the derivative with respect to $x$.
For the matter functions we obtain the equations
\begin{eqnarray}
(A N K')' = A K \left( \frac{K^2-1}{x^2} + H^2 - \frac{J^2}{A^2 N} 
 \right)
\ , \label{eqk} \end{eqnarray}
\begin{equation}
\left( \frac{x^2 J'}{A} \right)' = \frac{2 J K^2}{AN}
\ , \label{eqj} \end{equation}
and
\begin{eqnarray}
( x^2 A N H')' = A H \left( 2 K^2 + \beta^2 x^2 (H^2-1) \right)
\ . \label{eqh} \end{eqnarray}

Introducing the new function
\begin{equation}
 P = \frac{J}{A}
\label{p} \end{equation}
and using eq.~(\ref{eqa}), we see that
the metric function $A$ can be eliminated from the set of equations
(\ref{eqmu}), (\ref{eqk})-(\ref{eqh}).
These can then be solved separately, with the metric function $A$ 
being determined subsequently by (\ref{eqa}).

In the following we refer
to the above set of functions, variables and parameters,
also employed in \cite{bhk}, as notation I.

\subsection{Isotropic coordinates}

To make contact with the analytical results of \cite{bfm,bfm2}
and to extend them
to the case of gravitating dyons and dyonic black holes,
we now present the equations of motion in the notation of \cite{bfm2},
in the following referred to as notation II.
In the Schwarzschild gauge the following identifications hold
\begin{equation}
\begin{array}{ccccccccc}
\cite{bhk} : & x & A & N &
K & H & J & \alpha & \beta^2\\
\cite{bfm2} : & \alpha r & e^{\lambda+\nu} & e^{-2\lambda} &
W & \alpha H & J  & \alpha & \frac{\beta^2}{2}
\end{array}
\end{equation}

Along with \cite{bfm2}, we now employ
the gauge choice $e^{\lambda}={r(R)\over R}$, corresponding to
isotropic coordinates, and introduce the coordinate
\begin{equation}
\tau = \ln R
\ , \label{tau} \end{equation}
and the dot ( $\dot {} $ ) now indicates the derivative with respect to $\tau$.
To obtain a system of first order equations 
we further employ for the first derivatives the new variables \cite{bfm2}
\begin{equation}
    N={\dot r\over r}\ , \ \ \kappa=\dot \nu + N \ , 
\ \ U={\dot W\over r}\ , \ \ V=\dot H 
\ , \end{equation}
and we introduce
\begin{equation}
    B = e^{-\nu} J\ , \ \ C=\dot B
\ . \label{b} \end{equation}
(Note, that the symbol $N$ here
has a totally different meaning from eq.~(\ref{n}).)

Then the following system of 
first order autonomous equations is obtained
\begin{eqnarray}
\label{autoi}
\dot r &=& Nr\\
\dot N &=& N(\kappa-N)-(2U^2+V^2) - 2\alpha^2B^2W^2\\
\dot \kappa &=& 1-\kappa^2+2U^2-{\beta^2r^2\over 2} (H^2-\alpha^2)^2-2W^2H^2 + 2\alpha^2
W^2B^2\\
\dot H &=& V\\
\dot V &=& {\beta^2r^2\over 2} (H^2-\alpha^2)H+2W^2H- \kappa V\\
\dot W &=& rU\\
\dot U &=& U(N-\kappa)+ {W(W^2-1)\over r} + Wr (H^2-\alpha^2B^2)\\
\dot B &=& C\\
\dot C &=& -\kappa C+B(2W^2-4U^2-V^2+\kappa^2-1 \nonumber \\
       &\ & +2W^2H^2+{\beta^2r^2\over 2} (H^2-\alpha^2)^2) - 4\alpha^2B^3W^2
\label{autof}
\end{eqnarray}
supplemented by the constraint
\begin{equation}
\label{contr}
  2\kappa N = 1+N^2+2U^2+V^2-2V_2 
+ 2\alpha^2W^2B^2-\alpha^2(C+B(\kappa-N))^2
\end{equation}
compatible with the equations.

\section{Embedded Reissner-Nordstr\o m Solutions}

As noted long ago \cite{rn},
the Einstein-Yang-Mills-Higgs system admits embedded RN solutions.
In notation I, RN solutions
with mass $\mu_\infty$,
unit magnetic charge and arbitrary electric charge $Q$ are given by
\begin{equation}
\mu(x) = \mu_\infty - \frac{\alpha^2 (1 + Q^2)}{2x} , \ \ \ A(x)=1 
\ , \end{equation}
\begin{equation}
K(x)=0 \ , \ \ \
J(x) = 
  J_\infty
 - \frac{Q}{x} \ , \ \ \ 
H(x)=1 
\ , \label{Q} \end{equation}
where $J_\infty$ is a priori arbitrary, but here chosen as
$J_\infty = Q/x_h$ (see eq.~(\ref{Jhor})).
At the regular horizon $x_h$ the metric function 
\begin{equation}
  N(x) = {x^2 - 2x \mu_{\infty} + \alpha^2(1 + Q^2) \over x^2}
\end{equation}
vanishes, $N(x_h)=0$, yielding
\begin{equation}
      x_h = \mu_{\infty} + \sqrt{\mu_{\infty}^2 - \alpha^2 (1+Q^2)}
\end{equation}
or equivalently
\begin{equation}
     \mu_{\infty} = {x_h^2 + \alpha^2 (1+Q^2) \over 2 x_h}
\ . \end{equation}
For fixed values of $\alpha$ and $Q$, RN
solutions exist for 
$x_h \geq \alpha \sqrt{1+Q^2}$,
independent of $\beta$.

In the extremal case
\begin{equation}
 x_h =  \mu_{\infty} = \alpha \sqrt{1+Q^2}
\ , \label{exh} \end{equation}
extremal RN solutions are obtained.
Extremal RN solutions 
are characterized by $Q$ and $\alpha$.
In particular, for extremal RN solutions 
the function $B(x)$ defined in (\ref{b}) becomes constant
\begin{equation}
     B = {J \over A \sqrt{N}} = {Q \over \alpha \sqrt{1+Q^2}}
\ . \label{brn} \end{equation}

\section{Dyon Solutions}

Let us first consider the globally regular particle-like solutions
of the SU(2) EYMH system.

In the Prasad-Sommerfield limit, $\beta=0$,
the dyon solutions in flat space are known analytically \cite{jul},
whereas for finite $\beta$ they are obtained only numerically
\cite{bkt}.
In the presence of gravity, the corresponding gravitating 
monopole and dyon solutions are obtained only numerically as well.
The gravitating dyon solutions have many features in common
with the gravitating monopole solutions \cite{bhk}.

After presenting the boundary conditions for the asymptotically
flat globally regular solutions, we here briefly discuss
the fundamental dyon solutions
both in the Prasad-Sommerfield limit and for finite $\beta$. 
Then we turn to the excited dyon solutions.

\subsection{Boundary conditions}

Requiring asymptotically flat solutions implies in notation I
that the metric functions $A$ and $\mu$ both
approach a constant at infinity.
We here adopt
\begin{equation}
A(\infty)=1
\ , \end{equation}
and $\mu(\infty)=\mu_\infty$ represents the dimensionless mass
of the solutions.
The matter functions also approach constants asymptotically,
\begin{equation}
K(\infty)=0 \ , \ \ \ J(\infty)= J_\infty \ , \ \ \ H(\infty) = 1
\ , \end{equation}
where for magnetic monopole solutions $J_\infty = 0$.
The asymptotic fall-off of the function $J(x)$ determines the dimensionless
electric charge $Q$ (see eq.~(\ref{Q})).

Regularity of the solutions at the origin requires 
\begin{equation}
\mu(0)=0
\ , \end{equation}
and \cite{jul}
\begin{equation}
 K(0)=1 \ ,  \ \ \ J(0) = 0 \ , \ \ \ H(0)=0
\ . \label{bc0} \end{equation}
The local existence of a family of analytic solutions 
obeying these conditions is shown in
Appendix A.1.

\subsection{Fundamental dyons}

\boldmath
\indent {\bf $\beta=0$}
\unboldmath

Like the gravitating monopole solutions, the gravitating dyon
solutions exist up to
a maximal value of the coupling constant $\alpha$.
Beyond this value no dyon solutions exist.
The fundamental dyon branch 
does not end at the maximal value $\alpha_{\rm max}$,
but bends backwards and extends
up to the critical value $\alpha_c$.
Since $\alpha^2=4\pi G v^2$, variation of the coupling constant 
along the fundamental dyon branch can be considered
as first increasing $G$ with $v$ fixed 
up to the maximal value $\alpha_{\rm max}$,
and then decreasing $v$ with $G$ fixed 
up to the critical value $\alpha_c$.

Completely analogously to the fundamental monopole branch \cite{bfm},
the fundamental dyon branch reaches a limiting solution
at the critical value $\alpha_c$,
where it bifurcates with the branch of extremal RN solutions of
unit magnetic charge and electric charge $Q$ \cite{bhk}.
The critical value $\alpha_c(Q)$ depends only slightly on the charge $Q$
\cite{bhk}.

For $\alpha \rightarrow \alpha_c$
the minimum of the metric function $N(x)$ of the dyon solutions 
decreases monotonically, approaching zero
at $x_c = \alpha_c \sqrt{1+Q^2}$.
For $x \ge x_c$,
the metric function $N(x)$ of the limiting solution
corresponds to the metric function
of the extremal RN black hole with horizon
$x_h = x_c = \alpha_c \sqrt{1+Q^2}$,
unit magnetic charge and electric charge $Q$.
Likewise, for $x \ge x_c$,
the other functions of the limiting solution
correspond to those of this extremal RN black hole \cite{bhk}.
Consequently, also the mass of the limiting solution
coincides with the mass of this extremal RN black hole.

On the interval $0 \le x \le x_c$
the functions $N(x)$, $K(x)$ and $H(x)$ of the limiting solution
vary smoothly from their respective boundary values at $x=0$
to those of the extremal RN solution at $x_c$,
whereas the function $J(x)$ is identically zero.
The metric function $A(x)$ is identically zero, as well,
on the interval $0 \le x < x_c$,
but discontinuous at $x_c$.
Consequently, in the limiting spacetime
the inner and outer parts are not analytically connected
\cite{ewein,bfm}.

\boldmath
\indent {\bf $\beta>0$}
\unboldmath

Let us briefly consider the dyon solutions for
finite Higgs self-coupling, $\beta > 0$,
because we here encounter a phenomenon,
also present for the black hole solutions at 
vanishing Higgs self-coupling, $\beta=0$.

In flat space dyon solutions of arbitrarily large charge exist 
only for $\beta = 0$.
In contrast, for finite values of $\beta$ dyon solutions exist only up to
a maximal value of the charge, $Q_{\rm max}(\beta)$ \cite{bkt}.
As the maximal value of the charge is approached, 
the function $J(x)$ approaches asymptotically the limiting value
\begin{equation}
\lim_{Q \rightarrow Q_{\rm max}} J_\infty = 1 
\ . \label{Jlim} \end{equation}
At this point the solutions change character and become oscillating
instead of asymptotically decaying \cite{bkt}.

As illustrated in Figs.~1-2,
this phenomenon persists in curved space.
In Fig.~1 we show the mass
of the flat space dyon (solid line) and of the
gravitating dyon (dashed line, corresponding to $\alpha = 0.5$)
as a function of the charge $Q$
for $\beta^2=0$ and $\beta^2=0.5$. 
In Fig.~2 the corresponding asymptotic values of the function
$J(x)$ are presented.
We note, that for finite $\beta$ 
also gravitating dyon solutions exist only up to
a maximal value of the charge,
where $J_\infty \rightarrow 1$.

\subsection{Radially excited dyons}

Besides the branch of fundamental dyon solutions
there are branches of radially excited dyon solutions,
where the gauge field function $K(x)$ 
of the $n$-th radially excited dyon solution has $n$ nodes.

These solutions have no flat space counterparts, and
the variation of $\alpha$ along a branch of excited solutions
must be interpreted as a variation of $v$ with $G$ fixed.
In the limit $\alpha \rightarrow 0$ the Higgs field vacuum
expectation value therefore vanishes, while $G$ remains finite.
Because of the particular choice of dimensionless variables (\ref{xm}),
in this limit the solutions shrink to zero size
and their mass $\mu$ diverges.
As for the radially excited monopole solutions \cite{bfm},
the coordinate transformation $\tilde x = x/\alpha$
leads to finite limiting solutions in the limit $\alpha \rightarrow 0$,
which are the Bartnik-McKinnon solutions \cite{bhk}.

In Fig.~3 we show the normalized mass $\mu_\infty/\alpha$ 
as a function of $\alpha$
for the first excited dyon branch
with electric charge $Q=1$ and $\beta=0$.
For comparison also the
first excited monopole branch is shown.

Like the radially excited monopole solutions,
the radially excited dyon solutions 
exist only below a critical value 
of the coupling constant $\alpha$.
For the radially excited monopole solutions this critical value is 
$\alpha_c = \sqrt{3}/2$ \cite{bfm,bfm2}.
For the radially excited dyon solutions the critical value is larger 
and will be discussed in the context of the dyonic black hole solutions.

\section{Black Hole Solutions}

We now turn to the dyonic black hole solutions of the
SU(2) EYMH system, choosing $\beta=0$.
Again, the SU(2) EYMH dyonic black holes
have many features in common with the
magnetic non-abelian black holes \cite{bhk}.
In particular, 
dyonic non-abelian black hole solutions exist in a limited domain of the
$\alpha$-$x_h$ plane, which depends on the charge $Q$,
and, in the limit $x_h \rightarrow 0$,
the corresponding globally regular solutions are obtained.
But a new phenomenon occurs for small values of $\alpha$, 
where the non-abelian solutions no longer bifurcate with 
non-extremal RN solutions.

In the following, we briefly present the boundary conditions.
We generalize the considerations of \cite{bfm2} to dyonic solutions
and derive analytically the critical value 
of the coupling constant, $\hat \alpha$.
We then present our numerical calculations for the
fundamental dyonic black hole solutions and their radial excitations.

\subsection{Boundary conditions}

Imposing the condition of asymptotic flatness,
the black hole solutions satisfy the same
boundary conditions at infinity 
as the regular solutions.
The existence of a regular event horizon at $x_h$
leads in notation I to the conditions for the metric functions
\begin{equation}
\mu(x_h)= \frac{x_h}{2}
\ , \end{equation}
and $A(x_h) < \infty $,
and for the matter functions 
\begin{eqnarray}
 {N' K' }|_{x_h} =  \left. K \left( \frac{K^2-1}{x^2} + H^2 
 \right) \right|_{x_h}
\ , \end{eqnarray}
\begin{equation}
 J |_{x_h} =0
\ , \label{Jhor} \end{equation}
and
\begin{eqnarray}
 {x^2  N' H' }|_{x_h} =  \left. 
   H \left( 2 K^2 + \beta^2 x^2 (H^2-1) \right)
  \right|_{x_h}
\ . \end{eqnarray}
The local existence of a family of solutions 
obeying these conditions at the horizon is 
demonstrated in Appendix A.2.

\boldmath
\subsection{Critical value of $\alpha$}
\unboldmath

For magnetic black holes, there are two 
regimes of the coupling constant $\alpha$, 
corresponding to two distinct patterns of reaching 
a limiting solution as a function of the horizon radius $x_h$.
For larger values of $\alpha$, the solutions bifurcate with
an extremal RN black hole solution,
when the horizon radius appoaches its critical value $x_h^{\rm cr}$.
Here the solutions tend towards the corresponding extremal RN solution
with horizon radius $x_h^{\rm RN} = \alpha$
only on the interval $x \ge x_h^{\rm RN}$,
whereas they tend to a generic non-abelian limiting solution
on the interval $x_h^{\rm cr} \le x \le x_h^{\rm RN}$ \cite{bhk}.
In contrast for smaller values of $\alpha$ the solutions
bifurcate with a non-extremal RN solution on the full interval
$x \ge x_h^{\rm cr}$ \cite{bfm,bfm2}.
The transition point between these two regimes corresponds
to the critical value $\hat \alpha$, 
where the non-abelian black hole solutions tend towards the extremal RN
solution with horizon radius $x_h^{\rm RN} = \hat \alpha$
on the full interval $x \ge x_h^{\rm cr} = x_h^{\rm RN}$,
i.e.~$x_h^{\rm cr}$ and $x_h^{\rm RN}$ coincide.
For the magnetic non-abelian black holes
this critical value is given by
$\hat \alpha = \sqrt{3}/2$ \cite{bfm,bfm2}.

Analogously, for dyonic non-abelian black holes there are
also two regimes of the coupling constant $\alpha$, 
with $\hat \alpha$ marking
the transition point between these two regimes.
We now derive this critical value $\hat \alpha$
for the dyonic non-abelian black hole solutions, following closely
the reasoning of \cite{bfm2}.

Let us then consider the fixed points 
of the system (\ref{autoi})-(\ref{autof}). 
Employing notation II, the relevant fixed point is given by
the configuration
\begin{equation}
\label{fixp}
H_h = \alpha \ , \ \  W_h = 0 \ , \ \ \kappa_h=1 \ , \ \ 
r_h = {1\over  \sqrt{1- \alpha^2 B_h^2}}
\ , \end{equation}
where $N_h = V_h = U_h = C_h = 0$,
while $B_h$ is such that
$0 \leq B_h \leq 1/ \alpha$.

Linearization of the equations about this fixed point leads to the
following eigenvalues of the matrix defining the linear part 
(adopting the order $N$, $W$, $U$, $H$, $V$, $P$, $Q$, $\kappa$)
\begin{eqnarray}
\label{eig0}
\gamma_0 &=& 1\\
\gamma_{1,2} &=& -{1\over 2} \pm i \sqrt{   {3\over 4}-  \alpha^2
 {{1 -B^2_h} \over {1 - \alpha^2 B_h^2}}   }
\label{eig12}\\
\gamma_{3,4} &=& -{1\over 2} \pm \sqrt{{1\over 4} + \alpha^2\beta^2r_h^2}\\
\gamma_5 &=& 0\\
\gamma_6 &=& -1\\
\gamma_7 &=& -2
\label{eig7}
\end{eqnarray}
In particular,
we see that the eigenvalues $\gamma_{1,2}$,
which are related to the function $W$
and determine the critical value $\hat \alpha$ for the magnetic
black holes \cite{bfm2},
now depend on $B_h$. The zero mode $\gamma_5$ is related to the
occurrence of this free parameter.

For $Q=0$, the critical value of $\alpha$ corresponds to the
transition of the eigenvalue $\gamma_{1,2}$ (eq.~(\ref{eig12}) for $Q=0$)
from complex to real value \cite{bfm2}.
Repeating this reasoning for dyonic black holes,
we conclude, that the corresponding critical value of the coupling constant
for the dyonic non-abelian black holes is given by
\begin{equation}
\hat \alpha^2 = {3\over{4-B^2_h}}
\ . \label{acrit} \end{equation}

Here the relation of the critical value $\hat \alpha$ to 
the charge $Q$ seems only indirect, since the value of the function
$B(x)$ at the horizon enters in eq.~(\ref{acrit}).
However, for the dyonic non-abelian black holes
$\hat \alpha$ also represents the special point,
where the non-abelian black hole solutions tend towards the 
corresponding extremal RN
solution with horizon radius $x_h^{\rm RN} = \hat \alpha \sqrt{1+Q^2}$
on the full interval $x \ge x_h^{\rm cr} = x_h^{\rm RN}$.
This implies for the function $B(x)$
according to eq.~(\ref{brn})
\begin{equation}
\label{valp}
    \lim_{\alpha \rightarrow \hat \alpha} B(x) = 
{Q \over \hat \alpha \sqrt{1+Q^2} }  \ ,  \ \   \forall x > x_h^{\rm RN}
\ . \end{equation}
Consequently, we may insert for $B_h$ in eq.~(\ref{acrit})
the constant RN value, eq.~(\ref{brn}),
and obtain for the critical coupling constant the expression
\begin{equation}
    \hat \alpha(Q) =  \sqrt{{3 + 4 Q^2 \over 4(1+Q^2)}}
\ , \label{ahat} \end{equation}
which now depends on the electric charge $Q$.

\subsection{Fundamental dyonic black holes}

Let us now define some quantities helpful in the analysis
of the numerical results.
Denoting the set of functions of the extremal RN solution
with horizon radius $x_h^{\rm RN} = \alpha \sqrt{1+Q^2}$
by $(N_{\rm RN},K_{\rm RN},H_{\rm RN},P_{\rm RN})$,
and denoting the set of functions of the non-abelian black hole solution
with horizon radius $x_h$ (with $x_h < x_h^{\rm RN}$)
by $(N,K,H,P)$, we define 
\begin{equation}
     D N  = {\rm max}_{x \in [x_h^{\rm RN}, \infty]} 
  \vert N_{\rm RN}(x) - N(x) \vert 
\ , \label{dn} \end{equation}
and analogously we define the quantities $D K$, $D P$ and $D H$.
Clearly the quantities $DN$, $DK$, $DP$ and $DH$
must vanish, when the non-abelian black hole solutions
bifurcate with the corresponding extremal RN solution.
Therefore these quantities help
to determine the critical values with good accuracy 
in the regime $\alpha \ge \hat \alpha$.

\boldmath
{\rm $Q=0$}
\unboldmath

Our numerical evaluation of the domain of existence 
of the magnetic non-abelian black holes
in the $\alpha$-$x_h$ plane
is in agreement with the results of \cite{bfm,bfm2},
and in particular we confirm $\hat \alpha = \sqrt{3}/2$.
Considering the quantities $DN$, $DK$ and $DH$ (eq.~(\ref{dn}))
as functions of the horizon radius $x_h$,
we observe that for $\alpha \ge \hat \alpha$
the three functions approach zero at a common
value $x_h^{\rm cr}(\alpha) \le \hat \alpha$.
As $\alpha - \hat \alpha$ changes sign,
in particular the quantity $DK$ changes drastically,
becoming a convex function
when the critical horizon radius $x_h^{\rm cr}$
is approached.

\boldmath
{\rm $Q \ne 0$}
\unboldmath

We have numerically determined the domain of existence 
of the fundamental dyonic non-abelian black hole solutions
in the $\alpha$-$x_h$ plane,
which is shown in Fig.~4 for $Q=1$.
The straight line indicates the extremal RN solutions, dividing the
domain of existence into a lower region with non-abelian
black holes only and an upper region with both non-abelian and
non-extremal abelian black holes, completely analogously
to the case of magnetic black holes.

To start the detailed discussion of the diagram,
we choose a small fixed value of the horizon radius $x_h$
and vary $\alpha$.
Then the dyonic black hole solutions
exist only up to some maximal value $\alpha_{\rm max}(x_h)$.
>From $\alpha_{\rm max}(x_h)$
a second branch extends backwards until a critical
value $\alpha_c(x_h)$ is reached.
With increasing $x_h$ the second branch becomes increasingly smaller,
and finally disappears. 
The critical values $\alpha_{\rm max}(x_h)$ and
$\alpha_c(x_h)$ are both shown in Fig.~4.

The presence of the two branches is illustrated in Fig.~5 
for $Q=1$ and $x_h = 0.8$,
where the quantities $DN$, $DK$, $DP$ and $DH$
are shown as functions of the parameter $\alpha$.
Here the second branch of solutions only exists
in the small interval $1.1986 \le \alpha \le 1.2002$.

For fixed $x_h$ and varying $\alpha$,
the functions of the black holes solutions evolve
completely analogously to those of the regular dyon solutions.
Consequently, the limiting solution reached
at the critical value $\alpha_c(x_h)$
consists of two parts, the outer part
corresponding to the exterior of the extremal RN black hole with horizon
$x_h^{\rm RN} = \alpha_c \sqrt{1+Q^2}$,
unit magnetic charge and electric charge $Q$
and the inner part representing a genuine non-abelian solution.

Similarly for fixed $\alpha > \hat \alpha$ and varying $x_h$,
the limiting solution reached
consists of two parts, the outer part
corresponding to the exterior of the extremal RN black hole with horizon
$x_h^{\rm RN} = \alpha \sqrt{1+Q^2}$,
unit magnetic charge and electric charge $Q$,
as demonstrated in Figs.~6-7 for the black hole solutions
with horizon $x_h=1$, 1.1 and 1.153,
charge $Q=1$ and $\alpha=1$ for the functions
$N$ and $P$ and the functions $K$ and $H$, respectively.

According to eq.~(\ref{ahat}), for $Q=1$
the critical value $\hat \alpha$ is given by
$\hat \alpha = \sqrt{7/8} \approx 0.935$.
Here the transition
between the two regimes with different bifurcation patterns
is supposed to occur.
Indeed, our numerical calculations confirm this critical value,
as demonstrated in Figs.~8-9, where
the four quantities $DN$, $DK$, $DP$ and $DH$
are shown as functions of the horizon radius $x_h$
for $\alpha=0.94$ and $\alpha=0.93$, respectively.

For $\alpha=0.94$, which is slightly above $\hat \alpha$,
the curves in Fig.~8 clearly approach zero 
at $x_h^{\rm cr} \approx 1.235$. This value is well below 
the horizon value of the corresponding extremal RN solution,
$x_h^{\rm RN}= \alpha \sqrt{1+Q^2} \approx 1.329$.
Moreover, in this limit
$x_h \rightarrow x_h^{\rm cr}$,
the corresponding values of $K(x_h)$ and $H(x_h)$ 
do not tend to their RN values of zero and one, respectively,
but they converge to generic numbers in the interval $[0,1]$.

In contrast, for $\alpha=0.93$, which is slightly below $\hat \alpha$,
the function $DK$ becomes a convex function
when the critical horizon radius $x_h^{\rm cr}
\approx 0.93 \sqrt{2} \approx 1.315$ is approached.
Furthermore, the quantities $K(x_h)$ and $H(x_h)$ 
tend to their RN values zero and one, respectively,
for $x_h \rightarrow x_h^{\rm cr}$.
Thus for $\alpha = 0.93$ the limiting solution
represents a RN solution on the full interval
$x \ge x_h^{\rm cr}$.

Clearly, the transition point $\hat \alpha$
occurs inbetween these two values.
Approaching $\hat \alpha$ further from both sides, the value
$\hat \alpha = \sqrt{7/8} \approx 0.935$ is confirmed.
The gap in Fig.~4 on the rhs of $\hat \alpha$
represents a tiny region, numerically not accessible
with sufficient accuracy \cite{foot}.
We have confirmed the validity of eq.~(\ref{ahat})
also for other values of the charge $Q$.

In the regime $\alpha < \hat \alpha$, we observe that 
the maximal value of the horizon radius $x_h$ decreases
with decreasing $\alpha$,
completely analogously to the magnetic case.
Continuing our detailed discussion of the approach to the limiting
solution in this regime, however, we observe differences to
the magnetic case for smaller values of $\alpha$ and larger values of $Q$.

For fixed $\alpha$ close to $\hat \alpha$ and increasing $x_h$,
the dyonic black hole solutions bifurcate with a RN solution
when the maximal value of the horizon is reached.
The limiting solution corresponds to the exterior of the RN solution
on the full interval $x \ge x_h^{\rm cr}$.
For smaller values of $Q$ and
somewhat smaller fixed $\alpha$, with increasing $x_h$
the solutions do not yet bifurcate with a non-extremal RN solution,
when the maximal value of the horizon is reached.
Instead a second branch of solutions emerges,
extending backwards to lower values of the horizon radius,
until it bifurcates with a non-extremal RN solution
at the critical value of the horizon $x_h^{\rm cr}$ \cite{bfm,bfm2,bhk}.
Moving along both branches,
the values of $K(x_h)$ and $H(x_h)$ change monotonically,
reaching their respective RN values of zero and one as the critical
horizon is reached.

For the magnetic black holes as well as 
for the dyonic black holes with smaller values of $Q$
this latter bifurcation pattern persists
with decreasing $\alpha$ \cite{bfm,bfm2}.
In contrast,
for the dyonic non-abelian black holes with larger values of $Q$
new phenomena occur.

We exhibit these new phenomena in Fig.~10,
where the critical region of the $\alpha$-$x_h$ plane
shown in Fig.~4 for $Q=1$ is enlarged and more details are given.
First, we observe the occurrence of bifurcations,
which lead to the small triangular region $ABC$,
where three solutions exist.
Thus below $\hat \alpha$
we find four distinct regions in the $\alpha$-$x_h$ plane.
With decreasing $\alpha$ and
increasing $x_h$ there are
i) for $0.605 < \alpha$
first one, then zero solutions,
ii) for $0.575 < \alpha < 0.605$ 
first one, then three, then one, then zero solutions,
iii) for $0.535 < \alpha < 0.575$
first one, then three, then two, then zero solutions
and iv) for $\alpha < 0.535$
first one, then two, then zero solutions.

Second, with decreasing $\alpha$,
the value $J_\infty$ of the limiting solution increases.
When it reaches its maximal value of one,
the function $K(x)$ ceases to decay exponentially.
The limiting solution then no longer represents
a non-extremal RN solution,
instead an oscillating solution is reached as the limiting solution.
The corresponding critical point,
where the transition from the limiting non-extremal RN solutions
to the oscillating solutions occurs, is labelled $D$ in Fig.~10.
Along the curve $BD$
the maximal value of the horizon radius is reached 
when the limiting value $J_\infty=1$ is attained,
whereas along the curves $AB$ and $A'A$
an intermediate extremal value of the horizon radius is reached 
when the limiting value $J_\infty=1$ is attained.

Figs.~11-13 illustrate both phenomena,
representing the values $K(x_h)$, $H(x_h)$ and $J_\infty$
as functions of the horizon radius
for $\alpha=0.7$, 0.55 and 0.2, respectively.
As seen in Fig.~11, for $\alpha=0.7$
only one branch of solutions exists.
Here, with increasing $x_h$
$K(x_h)$ decreases monotonically to zero,
$H(x_h)$ increases monotonically to one,
while $J_\infty$ increases monotonically to a limiting value
smaller that one.
Decreasing $\alpha$, the critical point $C$ of Fig.~10 is reached,
below which three branches of solutions exist.
Moving along the three branches, 
$K(x_h)$ decreases monotonically, $H(x_h)$ increases monotonically
and $J_\infty$ increases monotonically until it reaches one,
as seen in Fig.~12 for $\alpha = 0.55$.
For still smaller values of $\alpha$, 
at the critical value $A$, the third branch
disappears, leaving two branches which end
when $J_\infty$ reaches one.
This is seen in Fig.~13 for $\alpha = 0.2$.
The maximal horizon radius in this case is approximately 0.917.  
For small values of $\alpha$ (i.e.~below the critical point $A$)
the lower curve in Fig.~4
represents the critical value of the horizon,
where the asymptotically exponentially decaying solutions cease to exist.

Considering the $Q$-dependence of the new phenomena, we observe that
the bifurcations leading to three branches arise only for $Q \ge 0.8$,
and the limiting value $J_\infty = 1$ is reached
only for $Q \ge 0.75$.
The latter is seen in Fig.~14, where
the critical value $D$ is shown, which marks the transition from
the limiting non-extremal RN solutions
to the oscillating solutions.
Also shown in Fig.~14 is the critical value $\hat \alpha$.
We note, that
with increasing $Q$ the two curves get increasingly closer.
  
\subsection{Excited dyonic black holes}

Besides the fundamental magnetic black hole solutions
there exist radially excited black hole solutions,
for which the gauge function $K(x)$ possesses $n$ nodes \cite{bfm,bfm2}.
These radially excited solutions
exist only for $\alpha \le \hat \alpha = \sqrt{3}/2$ \cite{bfm,bfm2}.

Our numerical analysis of the radially excited dyonic black hole solutions
strongly indicates that, 
at least for $n=1$, the excited dyonic black hole solutions penetrate 
into a small domain of the $\alpha > \hat \alpha$ regime,
as can already be anticipated from Fig.~3, where the
first excited regular dyon solution is shown for charge $Q=1$.
Indeed, we observe that the maximal value of $\alpha$, 
where the radially excited black hole solutions
cease to exist, increases more strongly with $Q$
than the critical value $\hat \alpha$.

Let us now consider the domain of existence of the black hole
solutions in the $\alpha-x_h$ plane in more detail.
In the regime $\alpha > \hat \alpha$,
for fixed $x_h$ and increasing $\alpha$
the solutions again reach a limiting solution
at the critical value $\alpha_c(x_h)$,
which consists of two parts, the outer part
corresponding to the exterior of the extremal RN black hole with horizon
$x_h^{\rm RN} = \alpha_c \sqrt{1+Q^2}$,
unit magnetic charge and electric charge $Q$.

For $\alpha \rightarrow \alpha_c(x_h)$
the function $N(x)$ develops a progressively decreasing minimum,
$N_{\rm min}$,
reaching zero at $x_h^{\rm RN}= \alpha_c(x_h) \sqrt{1+Q^2}$.
This is illustrated in Fig.~15, where the value of
$N_{\rm min}$ is shown as a function of $\alpha$ for the horizon radii
$x_h=0.01$, 0.4 and 0.8 and charge $Q=1$.
With increasing $x_h$ the value of $\alpha_c(x_h)$ decreases,
though it remains greater than $\hat \alpha$
even for $x_h=0.8$, as inspection of
the four quantities $DN$, $DK$, $DP$ and $DH$ reveals.

In Fig.~16 we show the functions $N(x)$, $K(x)$, $H(x)$ and $P(x)$
of the first radially excited black hole solution 
for charge $Q=1$, horizon radius $x_h=0.4$ and 
$\alpha=0.938 > \hat \alpha$.
The minimum of the function $N(x)$,
which here occurs close to $x_h^{\rm RN}= \alpha_c(x_h) \sqrt{1+Q^2}$,
remains well separated from the horizon $x_h$
for $\alpha \rightarrow \alpha_c(x_h)$.

Considering finally the first excited black hole solutions
for small fixed $\alpha$ and increasing $x_h$, we observe that
as for the fundamental black hole solutions,
the quantity $J_\infty$ here plays a major role.
With increasing $x_h$, $J_\infty$ approaches its maximal value one,
where the asymptotically exponentially decaying solutions cease to exist.
For $\alpha = 0.2$, for instance, this happens at $x_h \approx 0.28$.

\section{Conclusions}

In analogy to gravitating monopoles \cite{ewein,bfm,ewein3}
also gravitating dyons \cite{bhk} exist.
When a regular horizon $x_h$ is imposed,
from these solutions ``black holes within monopoles'' \cite{ewein,bfm}
and ``black holes within dyons'' \cite{bhk} emerge.
Besides the fundamental regular and black hole solutions
also radially excited solutions exist \cite{bfm,bfm2,bhk},
related to the corresponding EYM solutions \cite{bm,su2}.
The static spherically symmetric ``black holes within monopoles'' 
and ``black holes within dyons'' 
provide counterexamples to the ``no-hair conjecture''.

The domain of existence of the fundamental dyonic non-abelian black hole
solutions in the $\alpha-x_h$ plane is similar to the one of the
fundamental magnetic non-abelian black hole
solutions.
There are two regimes of the coupling constant $\alpha$.
For $\alpha < \hat \alpha$,
the maximal horizon radius increases with increasing $\alpha$, 
whereas for $\alpha > \hat \alpha$ it decreases.
The transition point between these two regimes, $\hat \alpha$,
depends on the charge $Q$ of the solutions,
$\hat \alpha =  \sqrt{(3 + 4 Q^2)/(4(1+Q^2))}$.

For fixed $\alpha$ and varying horizon radius,
a limiting solution is approached.
For $\alpha > \hat \alpha$ the limiting solution
consists of two distinct parts. The outer part
corresponds to the exterior of an extremal RN black hole solution
with horizon radius $x_h^{\rm RN} = \alpha \sqrt{1+Q^2}$,
unit magnetic charge and electric charge $Q$,
whereas on the interval $x_h \le x < x_h^{\rm RN}$
the limiting solution has the features of generic non-abelian solutions.
At the transition point $\alpha = \hat \alpha$ the limiting solution
corresponds precisely to the exterior of the extremal RN black hole solution
with horizon radius $x_h^{\rm RN} = \hat \alpha \sqrt{1+Q^2}$,
on the full interval $x \ge  x_h$.
In contrast, for $\alpha < \hat \alpha$
the limiting solution
represents the exterior of a non-extremal RN solution 
for the larger values of $\alpha$, while for the smaller values of
$\alpha$ the dyonic non-abelian black hole solutions
encounter a critical point, where the solutions
no longer are exponentially decaying but become oscillating.

The domain of existence of the 
radially excited dyonic non-abelian black hole
solutions in the $\alpha-x_h$ plane is similar to the one of the
fundamental solutions.
Again, there are two regimes of the coupling constant $\alpha$
with the transition point $\hat \alpha$.
This is in contrast to the
radially excited magnetic non-abelian black hole solutions,
which exist only up to $\hat \alpha$ \cite{bfm,bfm2}.

In a large part of their domain of existence,
the magnetic non-abelian black hole solutions are classically stable
\cite{ewein,bfm,bfm2,holl}.
In contrast, 
dyonic non-abelian black hole solutions should be classically unstable.
As for the classical dyon solutions in flat space,
their mass should be lowered continuously, 
by lowering the electric charge,
as long as there is no charge quantization
\cite{jul}.

\vfill
\newpage

\section{Appendix A}

We show the local existence of the non-abelian
dyon solutions at the singular point $x=0$ 
in Appendix A.1
and the local existence of the non-abelian
dyonic black hole solutions at the singular point $x_h$ 
in Appendix A.2.

\subsection{Appendix A.1}

As for the monopoles, the field equations for the dyons
have singular points at $x=0$ and $x=\infty$.
For the monopoles there exists a two-parameter family of local
solutions regular at $x=0$. Analogously, for the dyons
there exists a three-parameter family of local solutions
regular at $x=0$, as shown in the following.

Let us consider eqs.~(\ref{eqmu})-(\ref{eqh}), using notation I.
At the singular point $x=0$,
the expansion of regular solutions depends on four parameters, yielding
\begin{eqnarray}
\label{exp1}
N(x)&=& 1-n_2x^2+O(x^4) \ , \nonumber \\
A(x)&=& A_0 (1+a_2x^2+O(x^4)) \ , \nonumber \\
K(x)&=& 1-k_2x^2+O(x^4) \ , \\
J(x)&=& A_0(j_1x+O(x^3)) \ , \nonumber \\
H(x)&=& h_1x + O(x^3) \ , \nonumber 
\end{eqnarray}
where $A_0$, $h_1$, $j_1$ and $k_2$ are arbitrary parameters while
\begin{eqnarray}
a_2 &=& \alpha^2(j^2_1+4k^2_2+{h^2_1\over 2}) \ , \nonumber \\
n_2 &=& \alpha^2(j^2_1+h^2_1+4k^2_2 + {\beta^2\over 6})
\ . \end{eqnarray}

Eliminating the metric function $A$, 
we now show that the remaining equations for $N$, $K$, $H$ and $P$ 
admit a three-parameter family of local solutions near $x=0$,
analytic in the parameters.
This is the counterpart of Proposition 2 of Appendix A of \cite{bfm}
in the presence of the dyon degree of freedom.
Let us define
\begin{eqnarray}
& &
w_1 = {H\over x} \ , \ \  w_2 = NH' \  , \ \  
w_3 = {1-K \over {x^2}} \ , \ \  w_4 = {N K'\over x}
\ , \nonumber \\
& &
w_5 = {P\over x}\ , \ \ 
w_6 = (P'+PM)N \ , \ \ w_7 = {1-N\over {x^2}}
\   \end{eqnarray}
with 
\begin{equation}
    M = {A' \over A} = \alpha^2 x(
      {2 P^2 K^2 \over N^2 x^2} + {2 K'^2 \over x^2} + H'^2 )
\end{equation}
and furthermore
\begin{eqnarray}
& &
u_1 = {1\over 3}(2w_1+w_2) \ , \ \
u_2 = {1\over 3}(w_3-w_4) \ , \ \
u_3 = {1\over 3}(2w_5+w_6) \ , \nonumber \\
& &
v_1 = {1\over 3} (w_1-w_2) \ , \ \
v_2 = {1\over 3} (2w_3+w_4) \ , \ \
v_3 = {1\over 3} (w_5-w_6) \ , \nonumber \\
& &
v_4 = {w_7\over 2} - {\alpha^2\over 2} 
(u^2_1-2v^2_1+4 u^2_2 - 2v^2_2 + u^2_3
-2 v^2_3 + {\beta^2\over 6})
\ . \end{eqnarray}
Then eqs.~(\ref{eqmu}) and (\ref{eqk})-(\ref{eqh})
can be rewritten in the form
\begin{eqnarray}
\label{gen}
x u'_i &=& x^2f_i \ , \qquad \qquad i=1-3 \ , \nonumber\\
x v'_i &=& - \lambda_i v_i + x^2g_i \ \ , \qquad i=1-4
\ , \end{eqnarray}
where $f_i$ and g$_i$ are analytical functions of 
$x^2$, $u_j$, $v_j$ and $1/N$
and $\lambda_i = 3$, $i=1,2,3$. 
It now follows from Proposition 1 of Appendix A of \cite{bfm} 
that the system of equations
(\ref{eqmu}) and (\ref{eqk})-(\ref{eqh})
admits a three-parameter family of solutions of the form
\begin{equation} 
u_i = c_i + O(x^2) \ , \ \ v_i = O(x^2)
\end{equation}
which are locally analytic in $x^2$ and in the constants 
$c_i$. These constants correspond to 
$h_1$, $j_1$ and $k_2$ in (\ref{exp1}).

\subsection{Appendix A.2}

Following \cite{bfm2} and considering
the system of equations (\ref{autoi})-(\ref{autof}),
we here demonstrate the local existence of the non-abelian
dyonic black hole solutions at the singular point $x_h$.
Black hole solutions are characterized 
by a first order pole singularity of the function
$\kappa$ at the non-degenerate horizon \cite{bfm2}.
Due to translation invariance of the equations 
under $\tau \rightarrow \tau+\tau_0$,
we can assume that the horizon occurs at $\tau=0$. 

To rewrite the set of equations in the form (\ref{gen}), 
which reveals the number of
free parameters occurring in the expansion at the horizon
of the locally analytic solution,
we introduce the following set of functions
\begin{eqnarray}
& & (u_1, u_2 , u_3, u_4, u_5) = (r, W, H, 2\kappa N-2U^2-V^2, S) 
\ , \nonumber \\
& & (v_1, v_2, v_3, v_4) = 
 (\kappa-{1\over {\tau}}, U, V, Y-S(\kappa - {1\over {\tau}}))
\ , \end{eqnarray}
with 
\begin{equation}
S = B/\tau \ , \ \ Y=\dot S
\ . \end{equation}

These functions are regular at the horizon and can be expanded
in powers of $\tau$. 
Moreover, it follows that the equations for $u_i$ and $v_k$
are of the form (\ref{gen}) 
with $\lambda_1 = 2$, $\lambda_2 = \lambda_3 = 1$ and $\lambda_4 = 3$. 
This demonstrates the existence of a five-parameter family of local
analytic solutions of the equations. 
Because of the constraint (\ref{contr}), 
it actually reduces to a four-parameter family of local solutions.

\vfill
\newpage

{\bf Figure Captions}

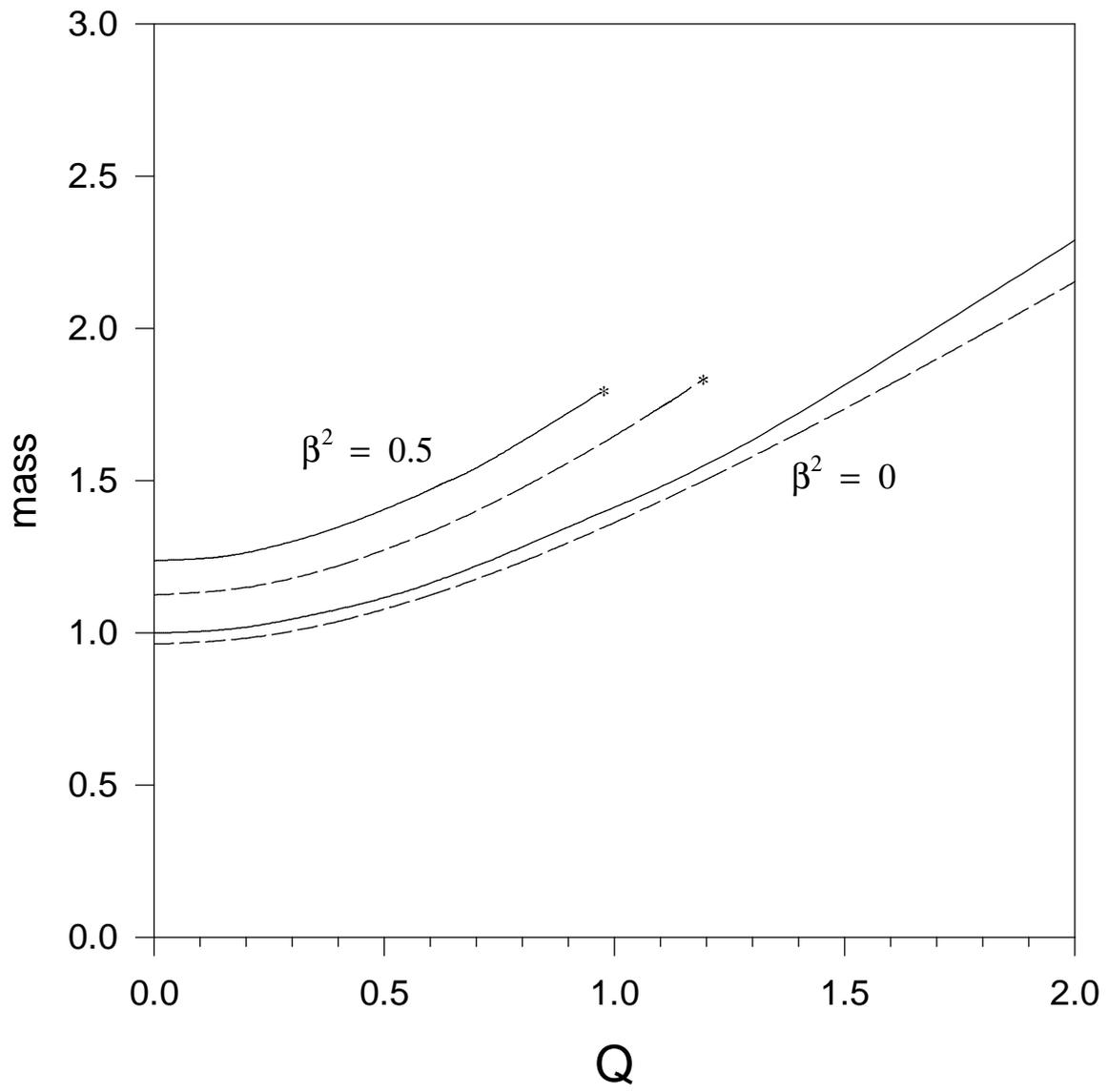
{\bf \noindent Figure 1}

The normalized mass of the regular dyon solutions is shown as a function
of the charge $Q$ for vanishing Higgs self-coupling ($\beta=0$)
and finite Higgs self-coupling ($\beta^2 =0.5$)
for the flat space solutions (solid lines) and curved space
solutions with $\alpha = 0.5$ (dashed lines).
The asterisks mark the transition from exponentially decaying
solutions to oscillating solutions.

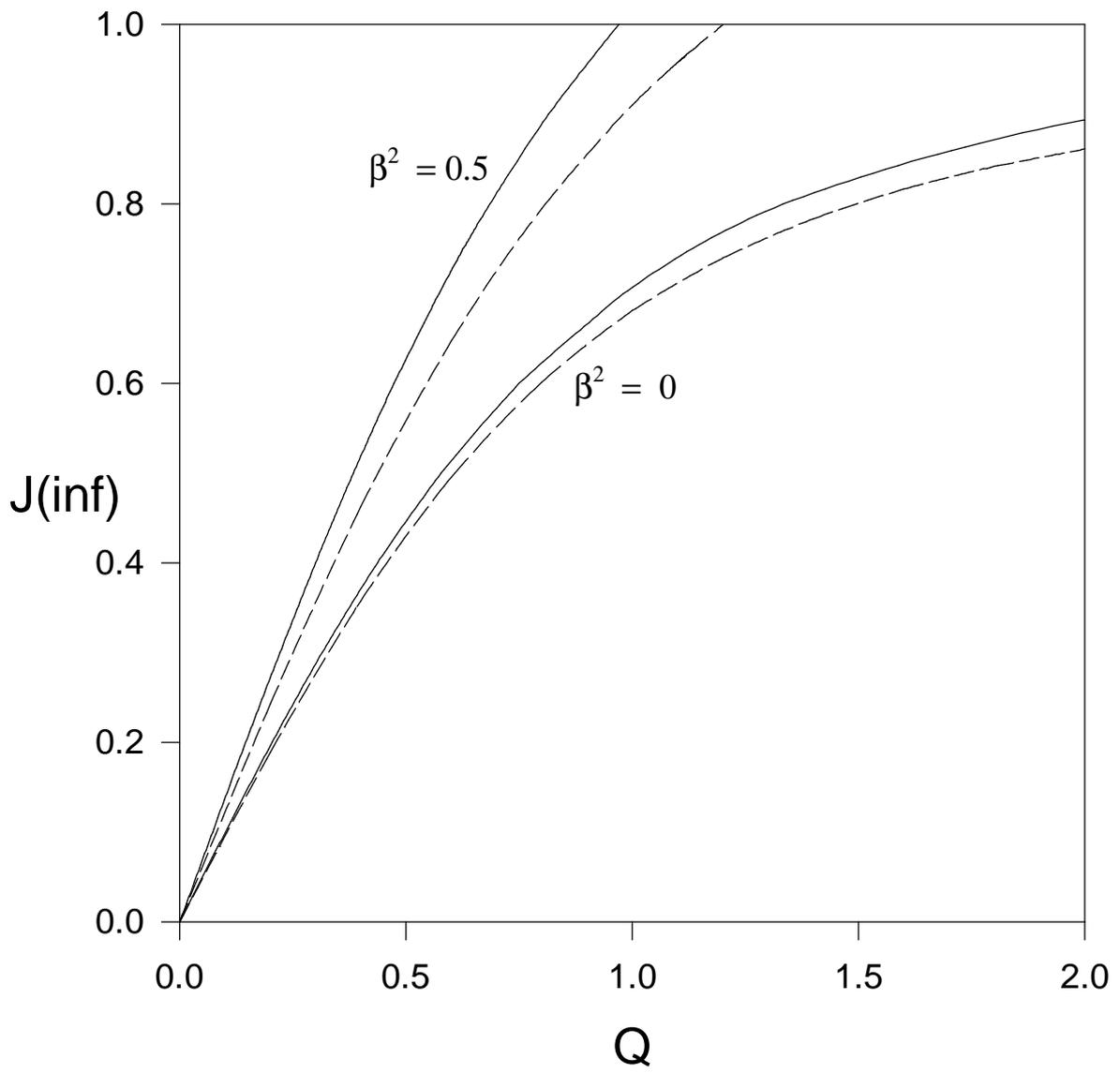
{\bf \noindent Figure 2}

Same as Fig.~1 for the asymptotic value of the function
$J(x)$.

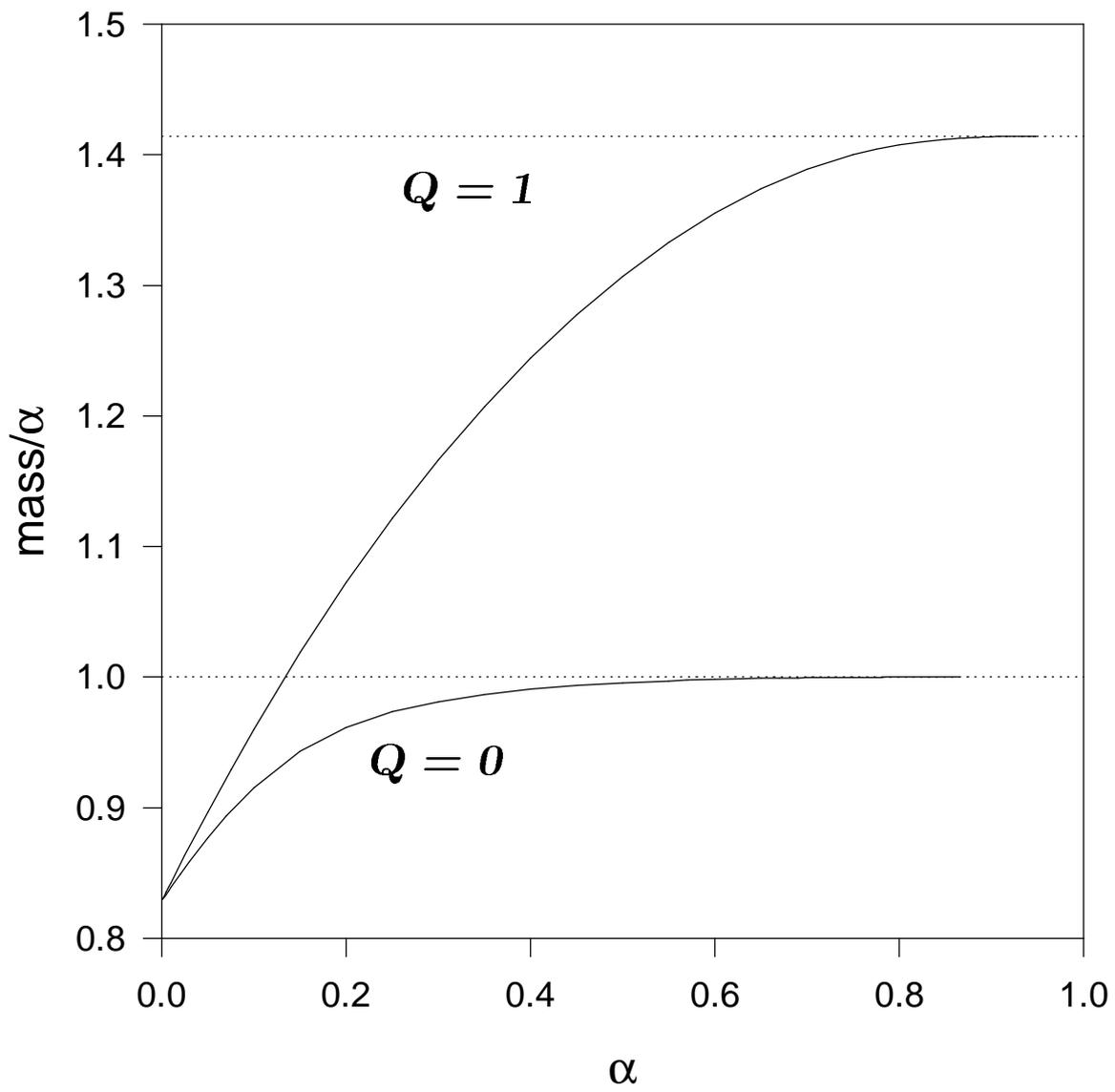
{\bf \noindent Figure 3}

The mass of the solutions of the 1st excited dyon branch for $Q=1$
and the 1st excited monopole branch ($Q=0$) 
is shown as a function of the coupling constant $\alpha$
for $\beta=0$ (solid).
Also shown is the mass of the branch of extremal RN solutions with
unit magnetic charge and $Q=1$ as well as $Q=0$ (dotted).
The normalization is chosen such that
at $\alpha=0$ the mass of the first Bartnik-McKinnon solution is obtained.

{\bf \noindent Figure 4}

The domain of existence of the dyonic black hole solutions
in the $\alpha-x_h$ plane is shown for $Q=1$.
The straight diagonal line represents the extremal RN solutions.
To the right of this line only non-abelian black hole solutions exist.
Here the limiting curve represents the maximal values of $\alpha$,
while the second curve close to it represents the critical values of
$\alpha$, where the solutions bifurcate with the
corresponding extremal RN solution.
The asterisk marks the critical value $\hat \alpha$.
To the left of the line of extremal RN solutions
both non-extremal RN solutions and non-abelian black hole solutions
coexist below the upper curve, representing the maximal
horizon radius as a function of $\alpha$. 
The small rectangle indicates the critical region,
where bifurcations occur (enlarged in Fig.~10).
Below the bifurcations,
the lower curve represents the critical values of the horizon,
where the solutions terminate because $J_\infty = 1$ is reached.

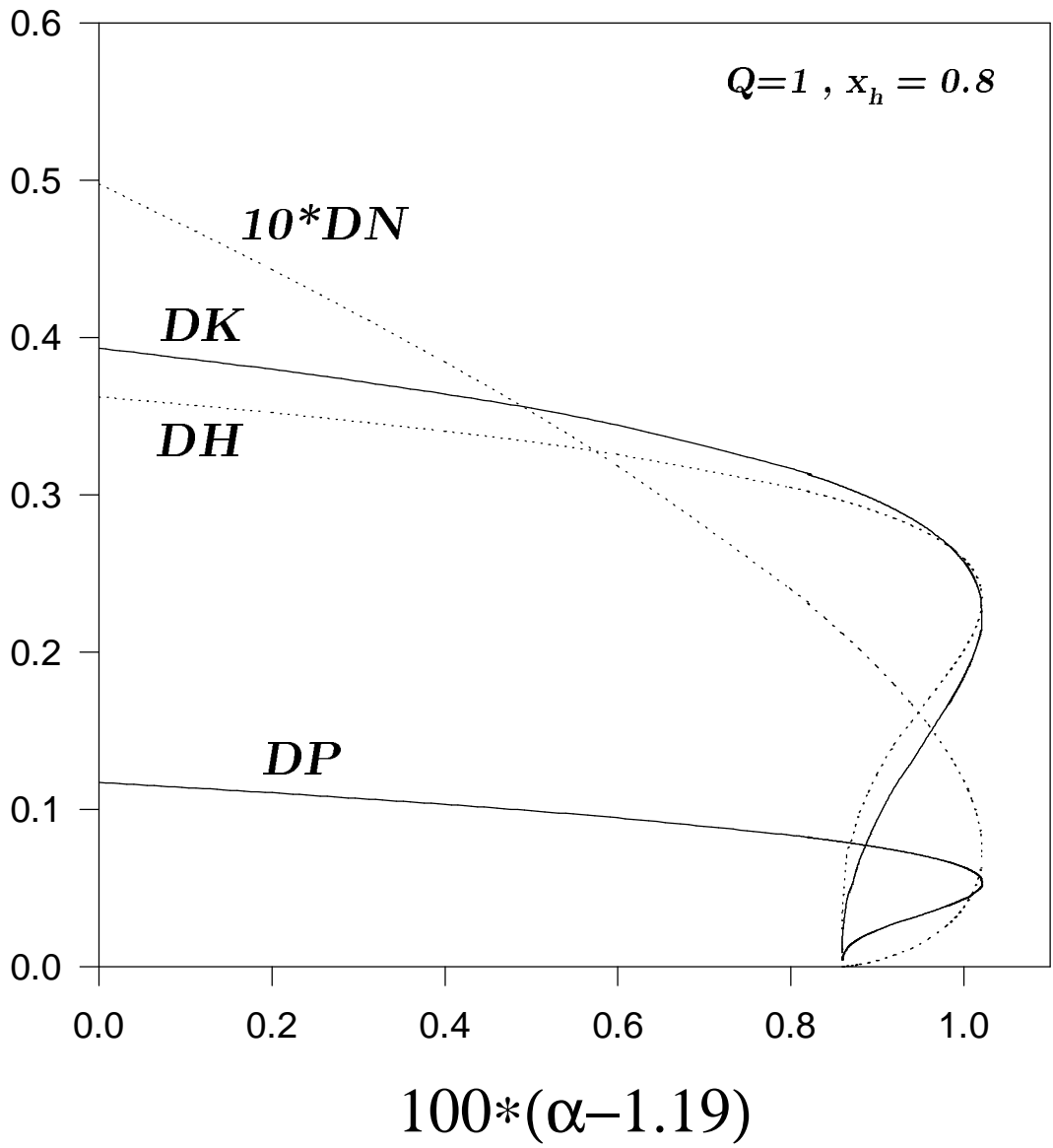
{\bf \noindent Figure 5}

The quantities $DN$, $DK$, $DP$ and $DH$
are shown as functions of the parameter $\alpha$
for the dyonic black hole solutions with
$Q=1$ and $x_h = 0.8$.
The second branch of solutions 
in the interval $1.1986 \le \alpha \le 1.2002$ is clearly visible.

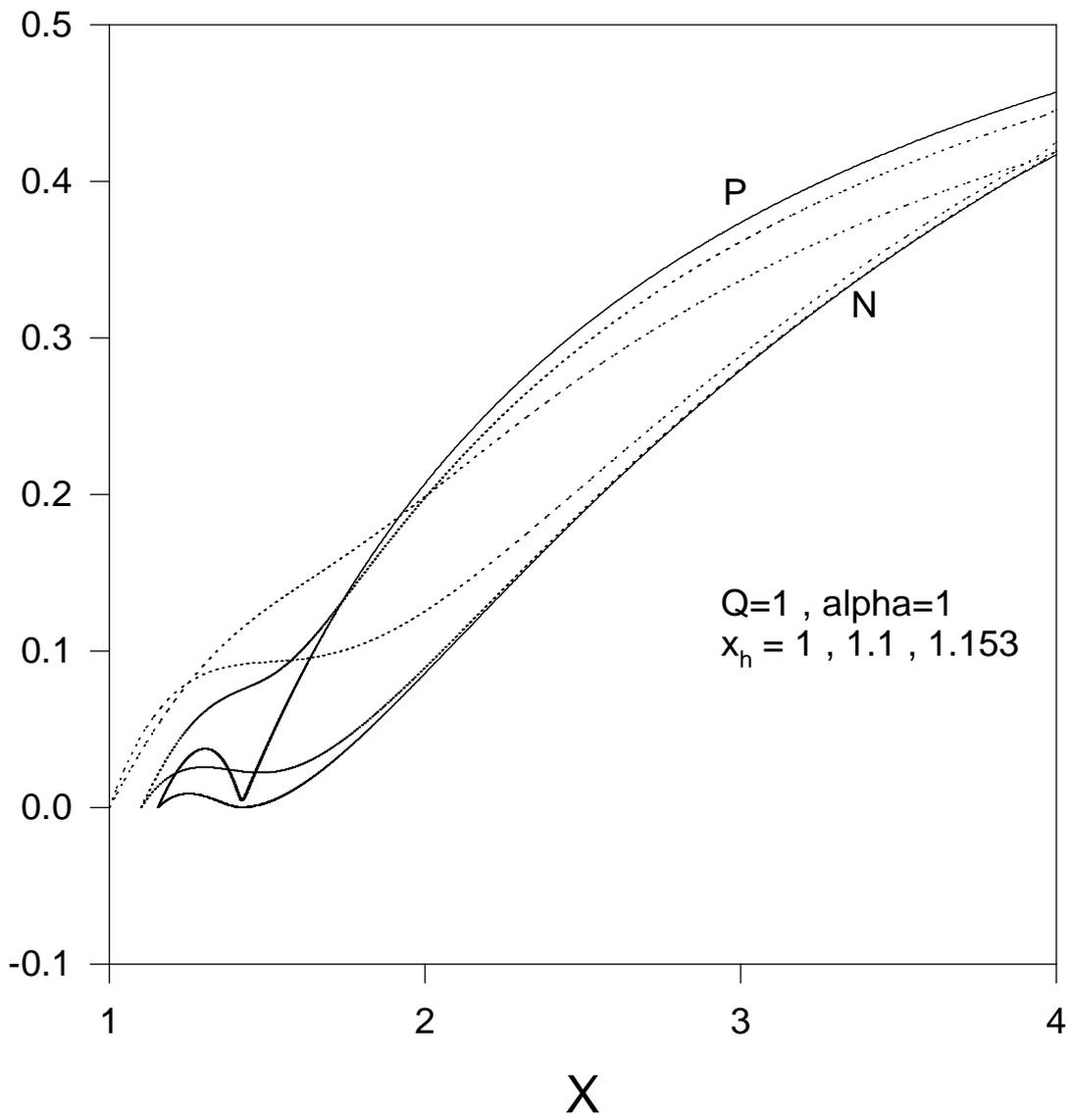
{\bf \noindent Figure 6}

The functions $N(x)$ and $P(x)$ are shown for the
dyonic black hole solutions with $Q=1$ and $\alpha=1$
for the horizon radii $x_h=1$, 1.1 and 1.153.

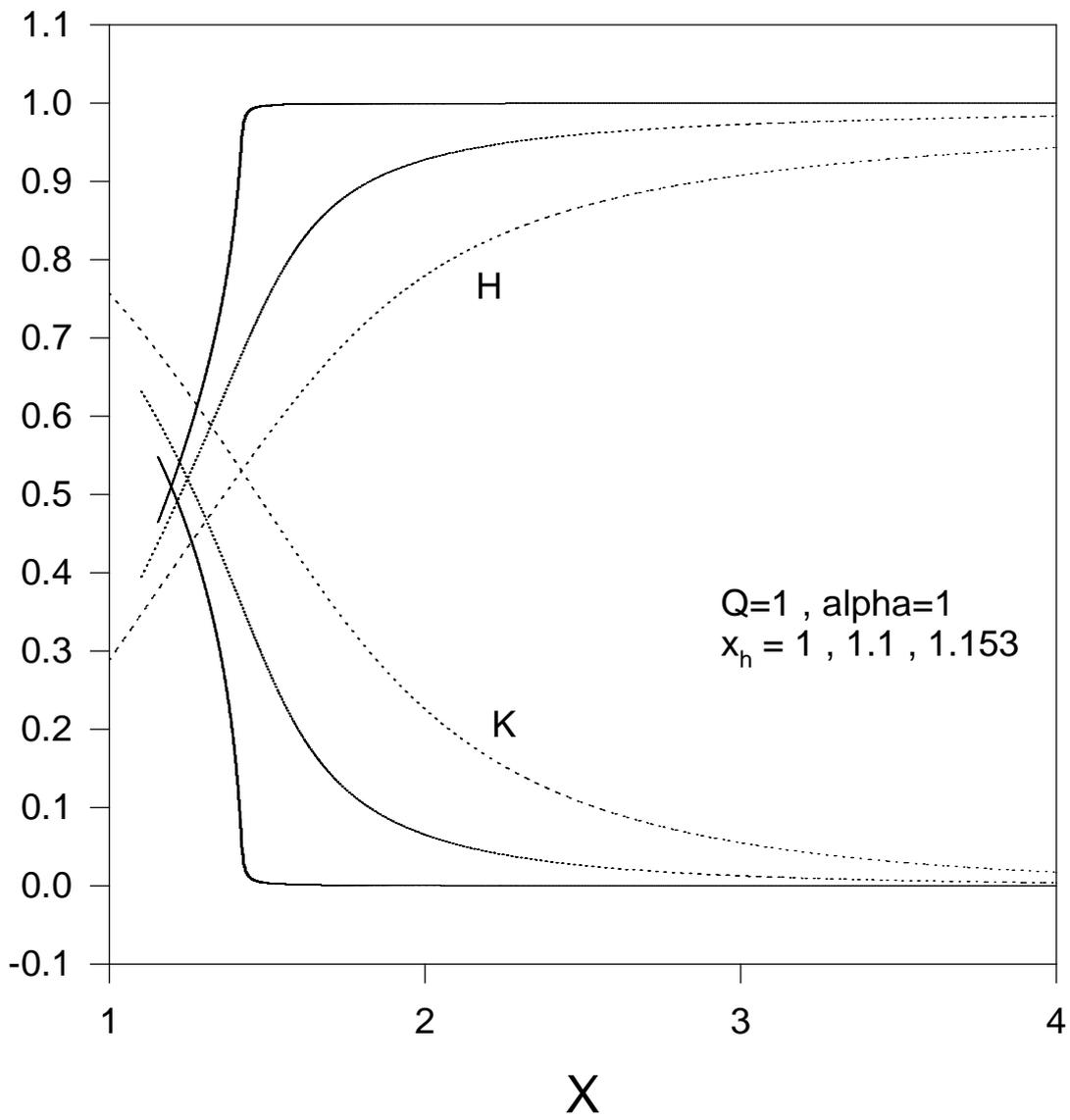
{\bf \noindent Figure 7}

Same as Fig.~6 for the functions $K(x)$ and $H(x)$.

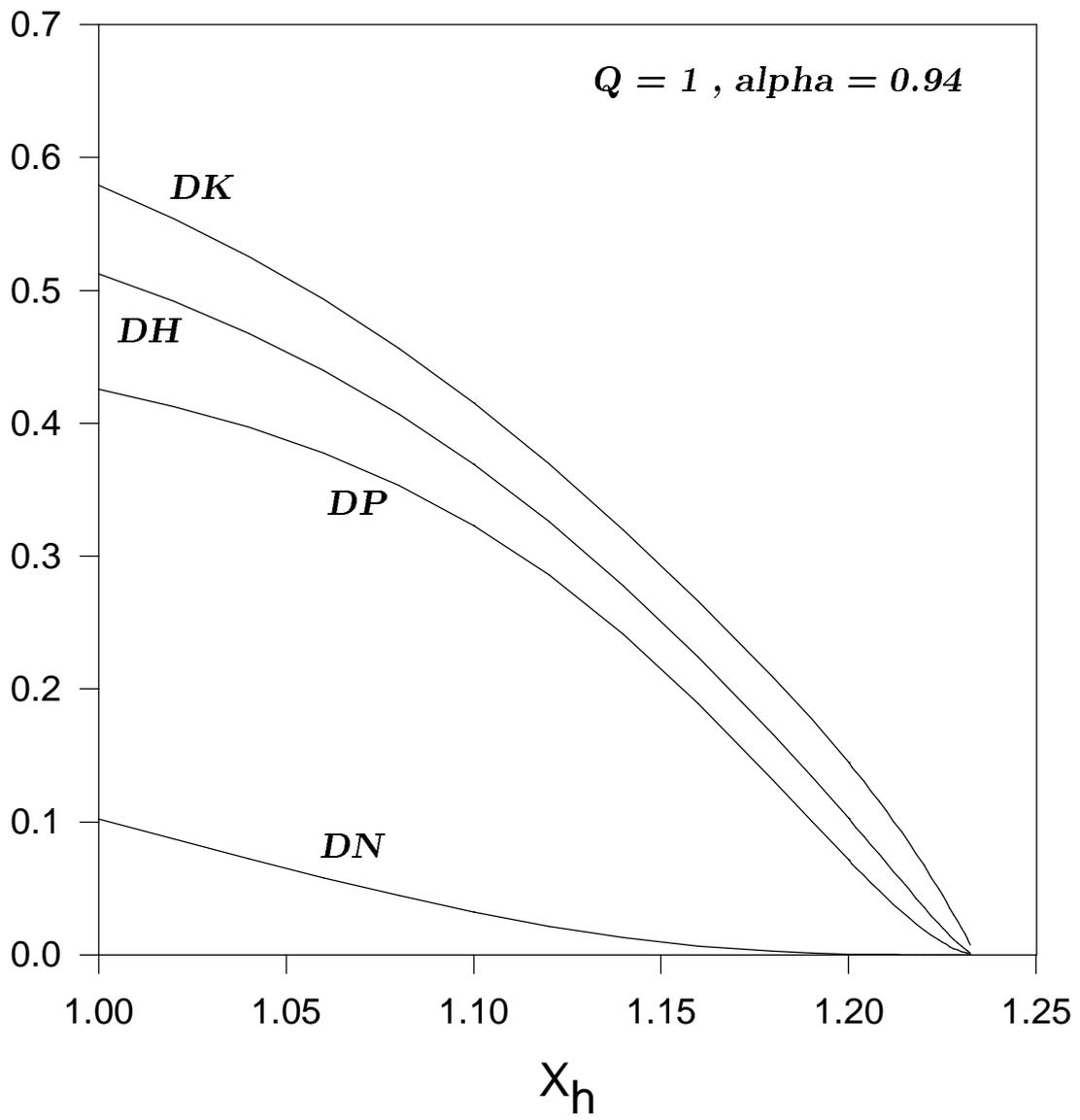
{\bf \noindent Figure 8}

The quantities $DN$, $DK$, $DP$ and $DH$
are shown as functions of the horizon radius $x_h$
for the dyonic black hole solutions with
$Q=1$ and $\alpha = 0.94 > \hat \alpha$.

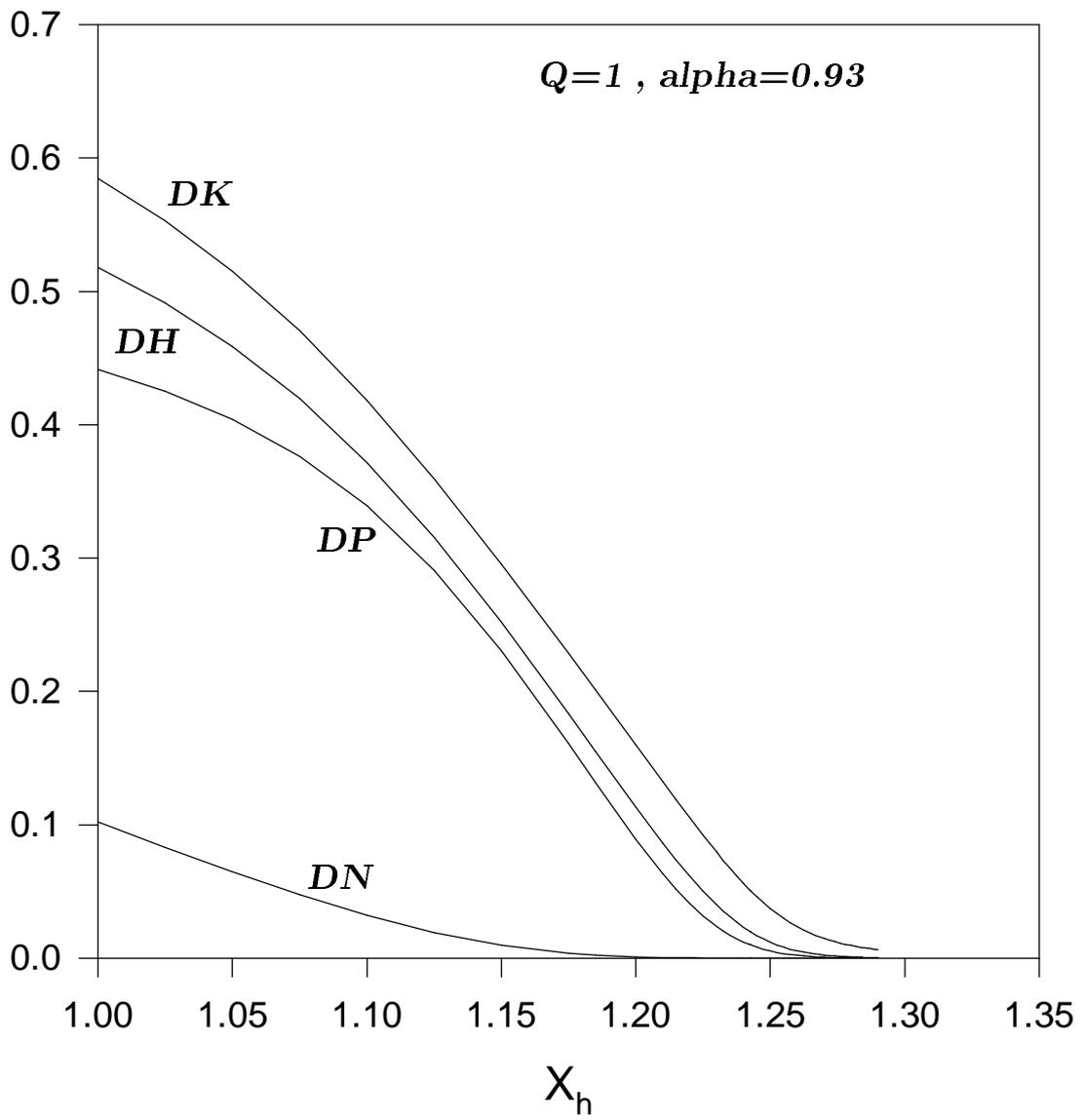
{\bf \noindent Figure 9}

Same as Fig.~8 for $\alpha = 0.93 < \hat \alpha$.

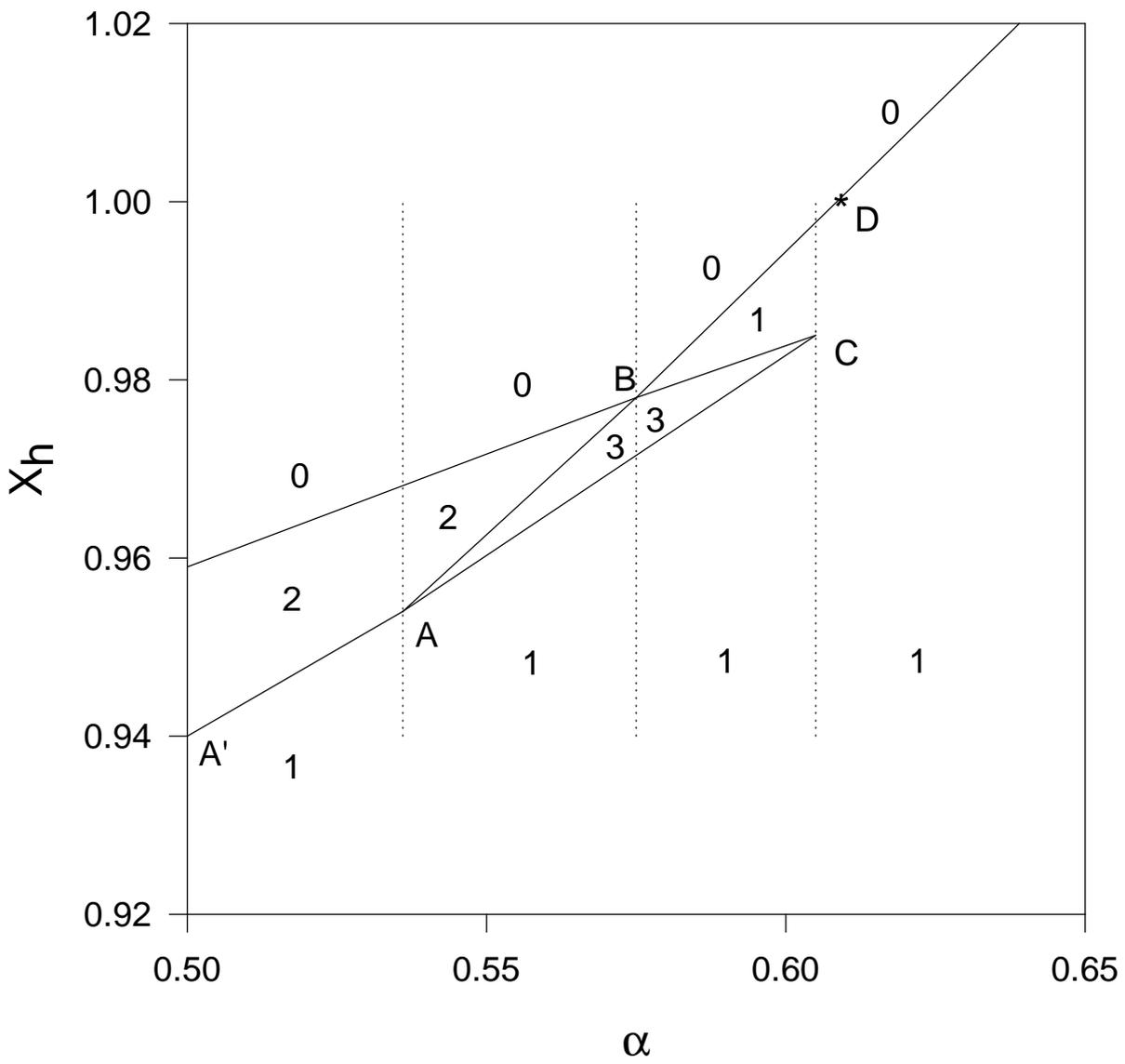
{\bf \noindent Figure 10}

Enlargement of the critical region
(small rectangle) of Fig.~4, showing
the domain of existence of the dyonic black hole solutions
in the $\alpha-x_h$ plane for $Q=1$.
The numbers 0-3 indicate the numbers of solutions
in the respective areas.
$A$ and $C$ indicate the critical values,
where the bifurcations occur,
$D$ indicates the critical value,
where the transition from the limiting non-extremal RN solutions
to the oscillating solutions occurs.
$J_\infty =1 $ is reached along the curve $A'ABD$.

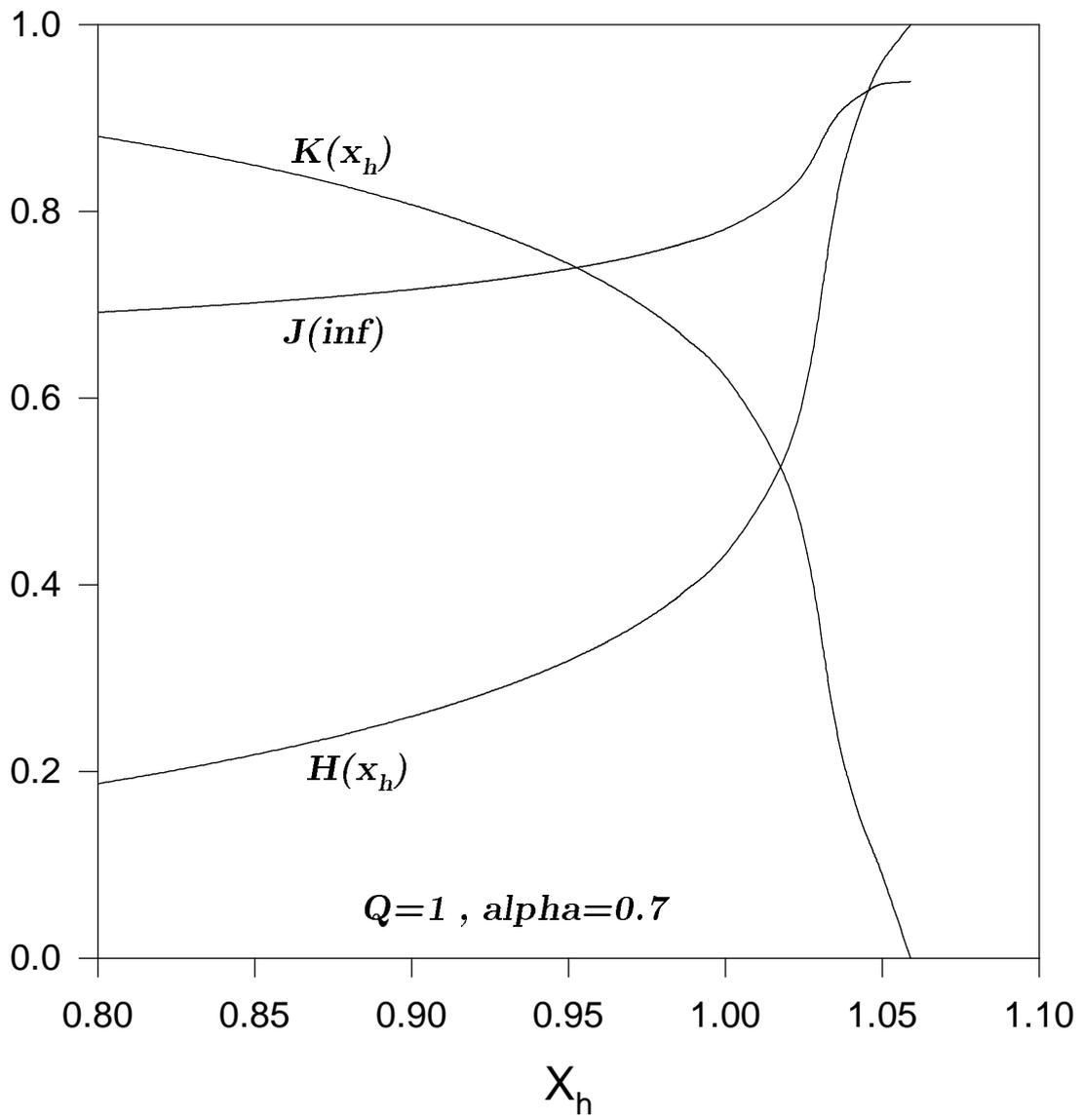
{\bf \noindent Figure 11}

The quantities $K(x_h)$, $H(x_h)$ and $J_\infty$
are shown as functions of the horizon radius $x_h$
for the dyonic black hole solutions with
$Q=1$ and $\alpha = 0.7$.

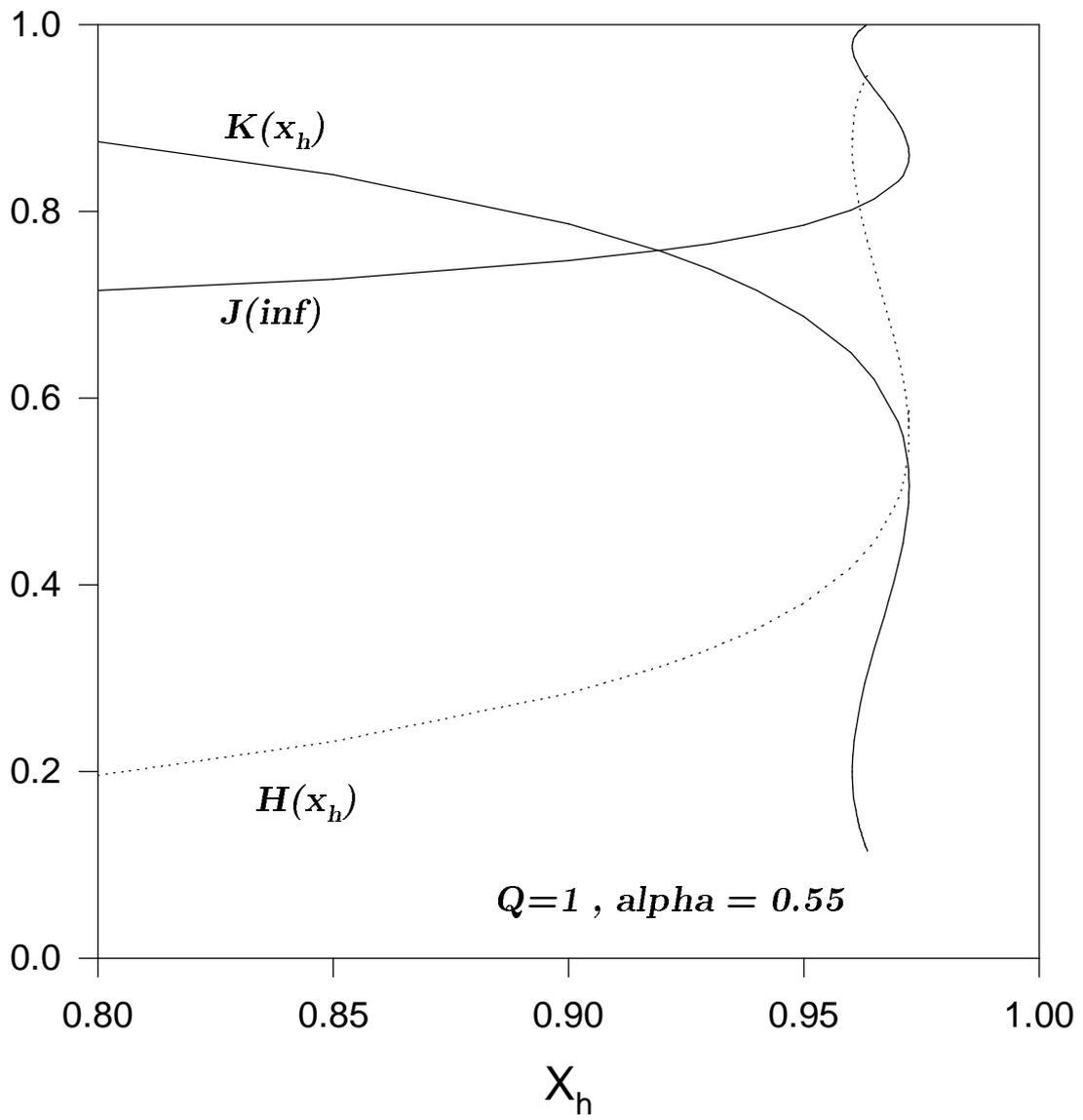
{\bf \noindent Figure 12}

Same as Fig.~11 for $\alpha = 0.55$.

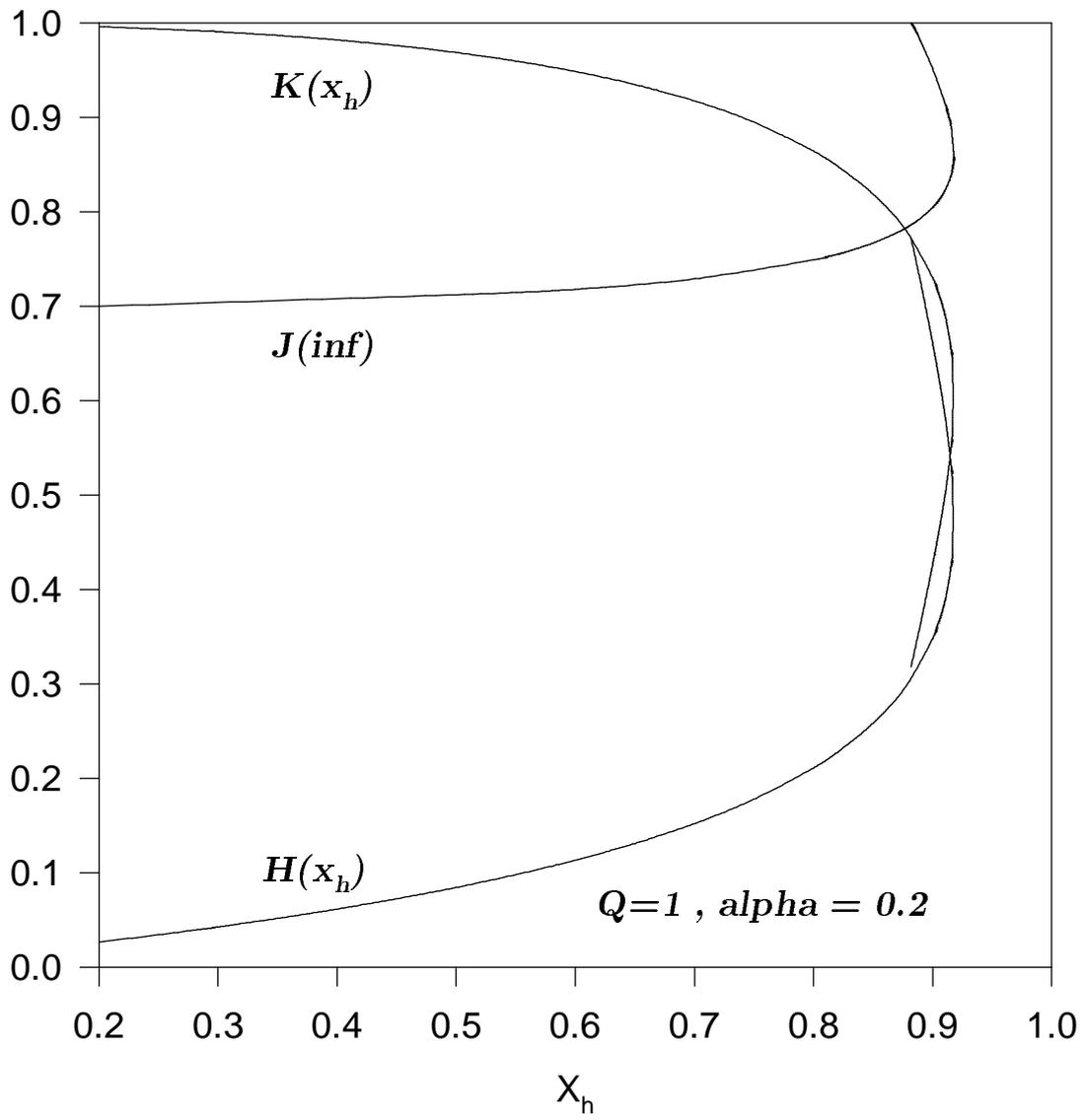
{\bf \noindent Figure 13}

Same as Fig.~11 for $\alpha = 0.2$.

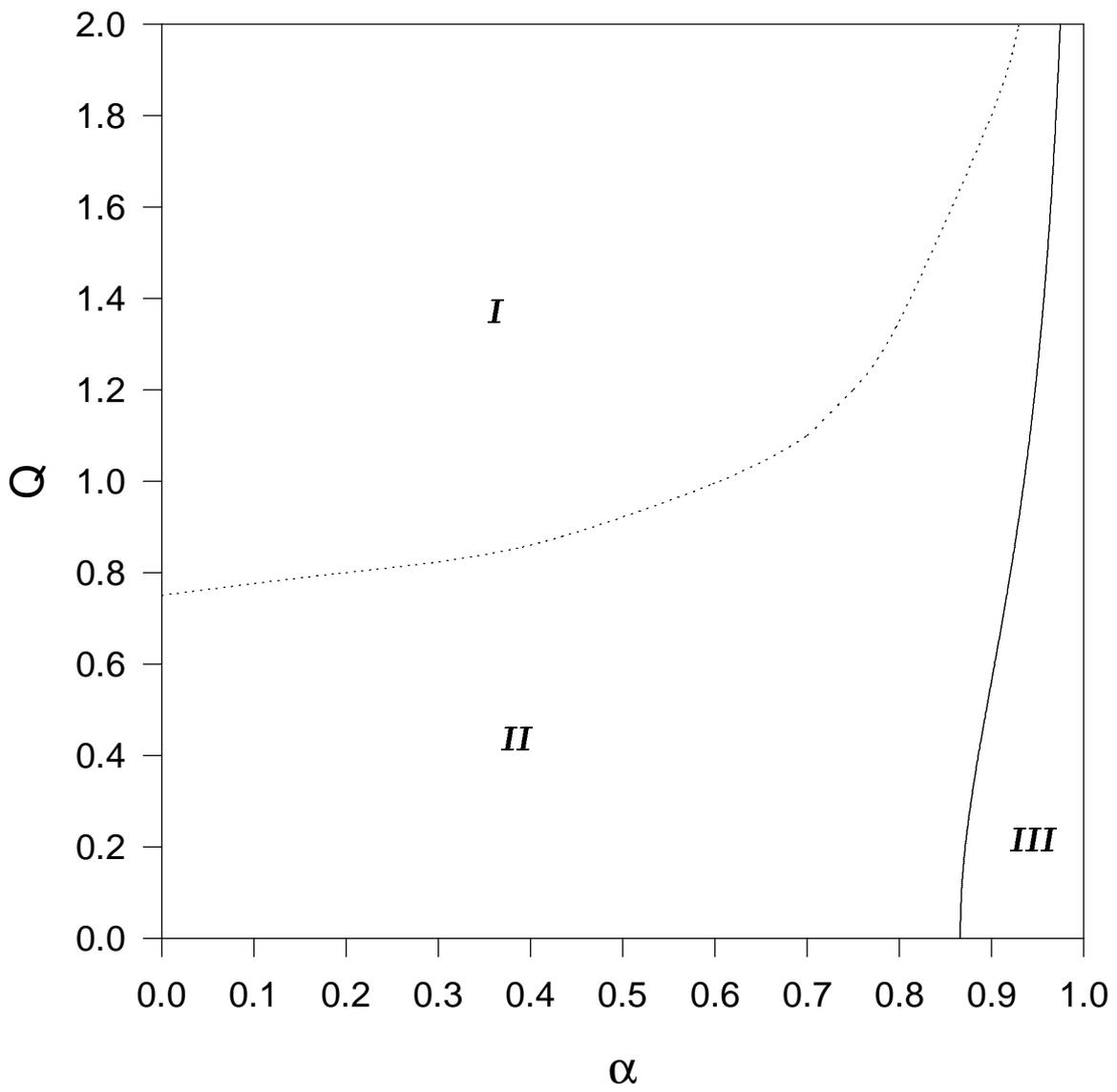
{\bf \noindent Figure 14}

The charge $Q$, corresponding to the critical value $D$,
where the transition from the limiting non-extremal RN solutions
to the oscillating solutions occurs,
is shown as a function of $\alpha$ (dotted).
Also shown is the critical value $\hat \alpha$
(solid).

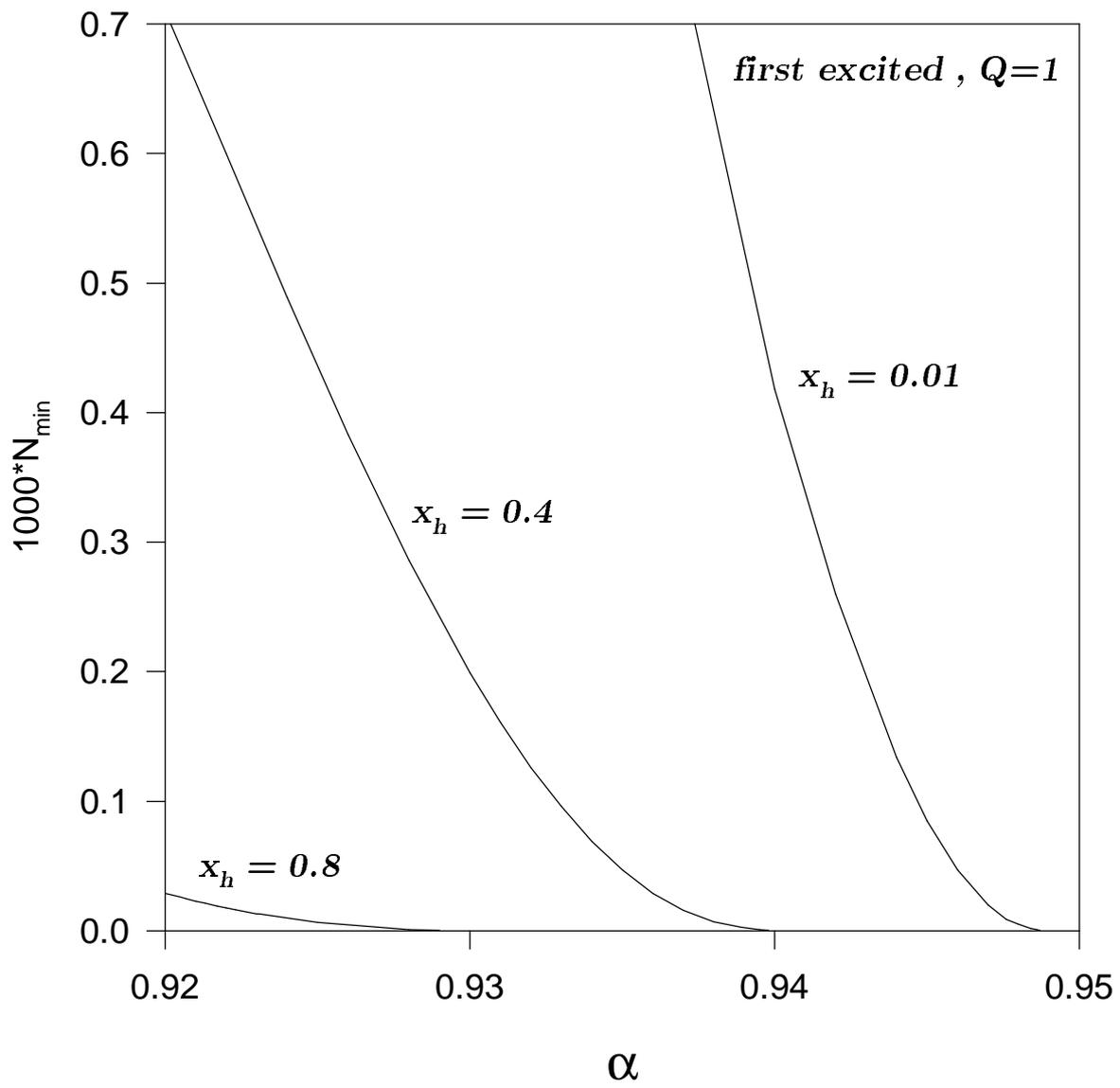
{\bf \noindent Figure 15}

The minimum of the function $N(x)$
is shown as a function of the parameter $\alpha$
for the first radially excited dyonic black hole solutions with
$Q=1$ and horizon radii $x_h = 0.01$, 0.4 and 0.8.

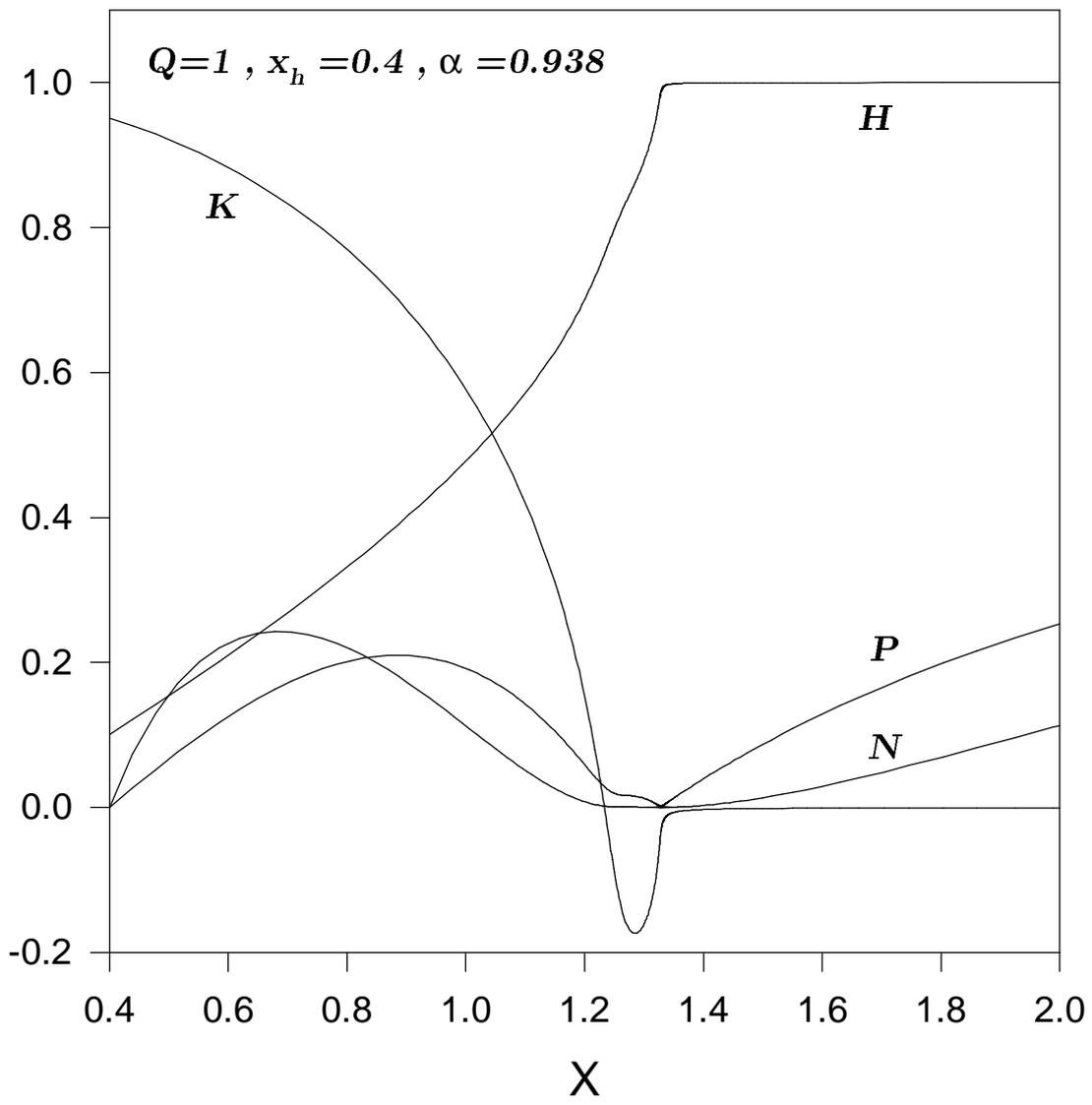
{\bf \noindent Figure 16}

The functions $N(x)$, $P(x)$, $K(x)$ and $H(x)$ are shown for the
first radially excited dyonic black hole solution with
$Q=1$, $\alpha=0.938$ and $x_h=0.4$.


\begin{thebibliography}{000}

\bibitem{thooft}
 G. 't Hooft,
 Nucl. Phys. B79 (1974) 276;\\
 A.M. Polyakov,
 JETP Lett. 20 (1974) 194.
\bibitem{jul}
 Poles with both magnetic and electric charges in non-Abelian
 gauge theory,
 Phys. Rev. D11 (1975) 2227;\\
 M.K. Prasad and C.M. Sommerfield,
 Exact classical solution for the 't Hooft Monopole and the
 Julia-Zee dyon,
 Phys. Rev. Lett. 35 (1975) 760.
\bibitem{old}
 P. van Nieuwenhuizen, D.Wilkinson and M.J.Perry,
 Phys. Rev. D13 (1976) 778.
\bibitem{ewein}
 K. Lee, V.P. Nair and E.J. Weinberg,
 Black holes in magnetic monopoles,
 Phys. Rev. D45 (1992) 2751.
\bibitem{bfm}
 P. Breitenlohner, P. Forgacs and D. Maison,
 Gravitating monopole solutions,
 Nucl. Phys. B383 (1992) 357.
\bibitem{bhk}
 Y. Brihaye, B. Hartmann, J. Kunz,
 Gravitating dyons and dyonic black holes,
 Phys. Lett. B441 (1998) 77.
\bibitem{ewein3}
 A. Lue and E.J. Weinberg,
 Magnetic monopoles near the black hole threshold,
 preprint hep-th/9905223.
\bibitem{bfm2}
 P. Breitenlohner, P. Forgacs and D. Maison,
 Gravitating monopole solutions II,
 Nucl. Phys. B442 (1995) 126.
\bibitem{aichel}
 P.C. Aichelburg and P. Bizon,
 Magnetically charged black holes and their stability,
 Phys. Rev. D48 (1993) 607.
\bibitem{rn}
 F.A. Bais and R.J. Russell, 
 Phys. Rev. D11 (1975) 2692;\\
 Y.M. Cho and P.G.O. Freund,
 Phys. Rev. D12 (1975) 1588.
\bibitem{gal1}
 A.A. Ershov and D.V. Gal'tsov,
 Phys. Lett. 150A (1990) 159.
\bibitem{gal2}
 D.V. Gal'tsov and A.A. Ershov,
 Non-abelian baldness of colored black holes,
 Phys. Lett. A138 (1989) 160.
\bibitem{popp}
 P. Bizon and O.T. Popp,
 No-hair theorem for spherical monopoles and dyons in SU(2)
 Einstein-Yang-Mills theory,
 Class. Quantum Grav. 9 (1992) 193.
\bibitem{bm}
 R. Bartnik, and J. McKinnon,
 Particlelike solutions of the Einstein-Yang-Mills equations,
 Phys. Rev. Lett. 61 (1988) 141.
\bibitem{bkt}
 Y. Brihaye, B. Kleihaus and D.H. Tchrakian,
 Dyon-Skyrmion lumps,
 preprint hep-th/9805059.
\bibitem{bck}
 P.~G. Bergmann, M. Cahen and A.~B. Komar,
 J. Math. Phys. 6 (1965) 1.
\bibitem{foot}
 For the magnetic non-abelian black holes a gap occurs
 at $\hat \alpha$ \cite{bfm2}.
\bibitem{su2}
 M.S. Volkov, and D.V. Gal'tsov,
 Black holes in Einstein-Yang-Mills theory,
 Sov. J. Nucl. Phys. 51 (1990) 747;\\
 P. Bizon, 
 Colored black holes,
 Phys. Rev. Lett. 64 (1990) 2844;\\
 H.~P. K\"unzle and A.~K.~M. Masoud-ul-Alam,
 Spherically symmetric static SU(2) Einstein-Yang-Mills fields,
 J. Math. Phys. 31 (1990) 928.
\bibitem{holl}
 H. Hollmann,
 On the Stability of Gravitating Nonabelian Monopoles,
 Phys. Lett. B338 (1994) 181.
 
\end{thebibliography}
\end{document}